\documentclass[letterpaper]{emulateapj}

\usepackage{epsfig}
\usepackage{graphicx}

\usepackage{multirow}

\usepackage{rotating}
\usepackage{natbib}
\usepackage{latexsym}
\bibpunct{(}{)}{;}{a}{}{,}

\begin{document}

\newcommand{\ledd}{%
$L_\mathrm{Edd}$}

\def\rem#1{{\bf #1}}
\def\hide#1{}

\newcommand{\specialcell}[2][c]{%
  \begin{tabular}[#1]{@{}c@{}}#2\end{tabular}}

\title{Millihertz quasi-periodic oscillations and thermonuclear bursts
from Terzan~5:\\ A showcase of burning regimes.}

\author{M. Linares\altaffilmark{1,5}, D. Altamirano\altaffilmark{2}, D. Chakrabarty\altaffilmark{1}, A. Cumming\altaffilmark{3}, L. Keek\altaffilmark{4}}

\submitted{Published in The Astrophysical Journal on 2012 March 9, Volume 748, Issue 2.}

\altaffiltext{1}{Massachusetts Institute of Technology - Kavli Institute for Astrophysics and Space Research, Cambridge, MA 02139, USA}

\altaffiltext{2}{Astronomical Institute ``Anton Pannekoek'', University of Amsterdam and Center for High-Energy Astrophysics, PO BOX 94249, 1090 GE, Amsterdam, Netherlands}

\altaffiltext{3}{Department of Physics, McGill University, 3600 rue University, Montreal, QC H3A 2T8, Canada}

\altaffiltext{4}{School of Physics and Astronomy, University of Minnesota, 116 Church Street SE, Minneapolis, MN 55455, USA}

\altaffiltext{5}{Rubicon Fellow}

\begin{abstract}

We present a comprehensive study of the thermonuclear bursts and
millihertz quasi-periodic oscillations (mHz QPOs) from the neutron
star (NS) transient and 11 Hz X-ray pulsar IGR~J17480--2446, located
in the globular cluster Terzan 5. The increase in burst rate that we
find during its 2010 outburst, when persistent luminosity rises from
0.1 to 0.5 times the Eddington limit, is in qualitative agreement with
thermonuclear burning theory yet opposite to all previous observations
of thermonuclear bursts. Thermonuclear bursts gradually evolved into a
mHz QPO when the accretion rate increased, and vice versa. The mHz
QPOs from IGR~J17480--2446 resemble those previously observed in other
accreting NSs, yet they feature lower frequencies (by a factor $\sim$3) and
occur when the persistent luminosity is higher (by a factor 4--25). We
find four distinct bursting regimes and a steep (close to inverse
cubic) decrease of the burst recurrence time with increasing
persistent luminosity. We compare these findings to nuclear burning
models and find evidence for a transition between the pure helium and
mixed hydrogen/helium ignition regimes when the persistent luminosity
was about 0.3 times the Eddington limit. We also point out important
discrepancies between the observed bursts and theory, which predicts
brighter and less frequent bursts, and suggest that an additional
source of heat in the NS envelope is required to reconcile the
observed and expected burst properties. We discuss the impact of NS
magnetic field and spin on the expected nuclear burning regimes, in
the context of this particular pulsar.

\end{abstract}


\maketitle

\section{Introduction}
\label{sec:intro}

%GENERAL INTRO THERMONUCLEAR BURSTS:

Matter accreted onto neutron stars (NSs) is piled up and compressed,
settling towards regions of increasing density and temperature. In
this process, depending on the rate at which accretion proceeds, both
stable and unstable thermonuclear burning of the accreted H and He
into heavier elements are expected \citep{Fujimoto81}. The main
parameter thought to determine the different burning regimes is the
mass accretion rate on the NS per unit surface area, $\dot{m}$
\citep[e.g.][]{Fujimoto81,Bildsten98}. When the burning layer becomes
thermally unstable heat cannot be transported as fast as it is
produced and a thermonuclear runaway occurs, producing a shell flash
that releases $10^{38}$-$10^{39}$ erg in tens of seconds \citep[we do
not discuss herein long bursts and superbursts, which are more
energetic, $10^{40}$-$10^{42}$ erg, and much less common;
e.g.,][]{Keek08}. Most of the energy is deposited in the outermost
layers of the NS within a few seconds and radiated away thermally for
tens of seconds while the photosphere cools down. Type I X-ray bursts,
which feature the spectral imprint of such photospheric cooling, were
discovered in low-mass X-ray binaries (LMXBs) more than 30 years ago
\citep{Grindlay76,Belian76,Hoffman78} and promptly identified as
thermonuclear bursts from accreting NSs
\citep{Woosley76,Maraschi77,Lewin77}. \citet{Joss80} pointed out that
a strong NS magnetic field can act to stabilize nuclear burning in
different ways, which may explain the fact that no thermonuclear
bursts have been observed to date from accreting NSs in high-mass
X-ray binaries (HMXBs).

%\rem{Andrew: brief intro on burning regimes?}

\begin{figure*}[t!]
\centering
%  \begin{center}
  \resizebox{2.1\columnwidth}{!}{\rotatebox{-90}{\includegraphics[]{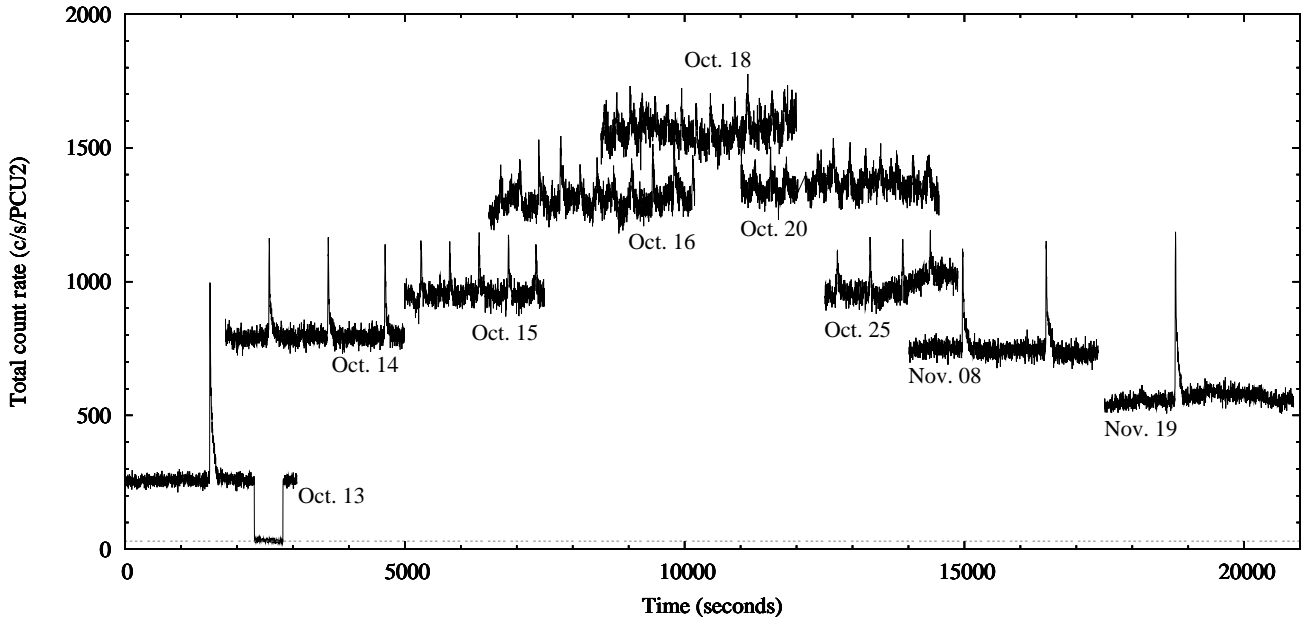}}}
  \caption{
Overview of the burst and persistent emission evolution along the
  outburst of T5X2. Black lines show total 2~s time resolution light
  curves during one {\it RXTE} orbit for nine selected dates, as
  indicated (using PCU2 and the full $\sim$2--60~keV band; gray dashed
  line shows the approximate, not subtracted, background rate). Times
  have been shifted arbitrarily for display purposes. The increase in
  burst rate and decrease in burst brightness as the persistent flux
  rises is evident. The step in October 13 was produced by a lunar
  eclipse \citep{Strohmayer10b}. The source became unobservable for
  {\it RXTE} after November 19 due to Solar constraints.
}
    \label{fig:evol}
\end{figure*}

%MARGINALLY STABLE BURNING AND MHZ QPOS:

Although direct observational evidence of stable thermonuclear burning
on accreting NSs has been elusive, as this is outshined by the much
more efficient accretion-powered ``persistent'' emission,
theoretical arguments and indirect observational evidence
\citep{Taam81,Fujimoto81,Paradijs88,Lewin93,Bildsten98}
suggest that at very high $\dot{m}$, close to or above the Eddington
mass accretion rate (Secs.~\ref{sec:data} \& \ref{sec:discussion}),
thermonuclear burning of the accreted H and He proceeds only stably.
Near the transition between unstable and stable burning, an
oscillatory burning regime was predicted by \citet{Paczynski83}, known
as marginally stable burning. \citet{Revnivtsev01} discovered
millihertz quasi-periodic oscillations (mHz QPOs) in the X-ray flux of
three atoll sources \citep[the sub-class of low luminosity
NS-LMXBs;][]{HK89}: 4U~1636-536, 4U~1608-52 and Aql~X-1 \citep[see
also][]{Strohmayer11}. They attributed this new phenomenon to
marginally stable burning on the NS surface. The mHz QPO frequency in
one of these systems has been found to decrease with time until a
bright type I X-ray burst occurs \citep{Altamirano08d}. The persistent
luminosity at which such mHz QPOs are observed has remained a puzzle,
as it suggests a critical $\dot{m}$ about an order of magnitude lower
than the stability boundary predicted by theory
\citep[e.g.,][]{Heger07}.

%T5X2 COMES INTO STAGE

On 2010 October 10, an X-ray transient in the direction of the
globular cluster Terzan~5 was discovered with the {\it International
Gamma-ray Astrophysics Laboratory}
\citep[][]{Bordas10,Chenevez10}. During the following week, {\it Rossi
X-ray Timing Explorer (RXTE)} observations revealed 11~Hz pulsations
\citep{Strohmayer10} and burst oscillations at the same frequency
\citep{Altamirano10a,Cavecchi11}. The {\it Chandra} localization
\citep{Pooley10} confirmed that this was a new NS transient, named
IGR~J17480--2446 \citep[labeled CX25 or CXOGlb J174804.8--244648
by][]{Heinke06}. We refer hereinafter to IGR~J17480--2446 as T5X2, as
this is the second bright X-ray source discovered in Terzan~5 (after
EXO~1745--248). A 21.3~hr orbital period was measured from the Doppler
shifts on the pulsar frequency \citep{Strohmayer10b,Papitto11}, and
the NS magnetic field was estimated to be between
10$^{8}$--10$^{10}$~G based on the inferred magnetospheric radius
\citep{Papitto11,Miller11}. This makes T5X2 the type I X-ray burst
source (burster) with the slowest known NS spin and with the highest
estimates of the NS magnetic field strength. Near the outburst peak
T5X2 showed X-ray spectral and variability behavior typical of Z
sources \citep[the sub-class of high luminosity NS-LMXBs;][]{HK89},
when it was accreting at about half of the Eddington rate
\citep{Altamirano10b}.

\citet{Linares10c} argued that all the X-ray bursts from T5X2 had a
thermonuclear origin, based on the evolution of the burst rate. Given
the lack of spectral softening along the tail of many of the bursts
and their short recurrence times, \citet{Galloway10} suggested that
some of the T5X2 bursts were type~II instead of type~I (i.e.,
accretion- instead of nuclear-powered). However, the
persistent-to-burst energy ratio throughout the October--November
outburst of T5X2 was typical of type~I X-ray bursts, i.e., fully
consistent with the accretion-to-thermonuclear efficiency ratio
\citep{Linares11b,Chakraborty11,Motta11}. Furthermore,
\citet{Linares11b} measured a smooth evolution of the burst luminosity
and spectral profiles and put forward a scenario to explain the lack
of cooling in the faintest bursts, conclusively identifying all X-ray
bursts detected from T5X2 as thermonuclear.

%THIS WORK:

We present a thorough analysis of the mHz QPOs from T5X2, including
but not limited to the ones originally reported by
\citet{Linares10c}. We study the mHz QPO frequency evolution and
energy-dependent amplitude, as well as all X-ray bursts from T5X2
detected with {\it RXTE} while the persistent (accretion) luminosity
varied along the outburst. Unlike previous studies
\citep{Motta11,Chakraborty11}, we analyze the complete sample of {\it
RXTE} bursts and compare their properties to theoretical models of
thermonuclear burning, along the full range in persistent luminosity
($\sim$10--50\% of the Eddington luminosity). Section~\ref{sec:data}
gives the details of the data analysis, and Section~\ref{sec:results}
presents the main observational results: a smooth evolution between
bursts and mHz QPOs (Figure~\ref{fig:evol}), the mHz QPO properties in
detail and four different bursting regimes during the October-November
2010 outburst of T5X2. In Section~\ref{sec:discussion} we place the
unique mHz QPO and bursting behavior of T5X2 in the framework of
thermonuclear burning theory and discuss the possible effects of
composition, NS spin and magnetic field on the observed bursting
properties. Section \ref{sec:conclusions} gives our summary and
conclusions.

\section{Data analysis}
\label{sec:data}

We analyzed all {\it RXTE} observations of T5X2 during its
October--December 2010 outburst: a total of 46 observations taken
between 2010 October 13 and 2010 November 19 (proposal-target number
95437-01). The source became Sun constrained after that date, and was
not detected by the {\it Monitor of All-sky X-ray Image (MAXI)} on
2010 December 28, indicating that the outburst finished between 2010
November 19 and 2010 December 28.
We visually searched for X-ray bursts in the full dataset, using 2~s
time resolution 2--30~keV lightcurves. We performed time-resolved
spectroscopy of all bursts using high time resolution data
(E\_125us\_64M\_0\_1s, or GoodXenon when available). We extracted
dead-time corrected spectra in 2~s time bins, using a $\sim$100~s-long
pre- or post-burst interval as background. We added a 1\% systematic
error to all channels, grouped them to a minimum of 15 counts per
channel when necessary and fitted the resulting spectra within Xspec
(v 12.6.0q), using a simple blackbody model with the absorbing column
density fixed to 1.2$\times$10$^{22}$ cm$^{-2}$ \citep{Heinke06}. We
used a distance to T5X2 of 6.3~kpc, the highest value reported from
{\it HST} photometry of Terzan~5 \citep{Ortolani07}, consistent with
the distance measurement based on photospheric radius expansion bursts
from another burster in the same globular cluster
\citep{Galloway08}. We note, however, that recent estimates of this
distance range between 4.6~kpc and 8.7~kpc
\citep{Cohn02,Ortolani07,Lanzoni10}, and therefore any value of the
luminosity, energy and mass accretion rate has a systematic
uncertainty of a factor $\sim$3.6.

We measured the burst rise time \citep[$t_\mathrm{rise}$, as defined in
][i.e., the time to go from 25\% to 90\% of the peak count
rate]{Galloway08}, and the total radiated energy ($E_\mathrm{b}$) by
integrating the bolometric luminosity along each burst. We defined the
wait time, $t_\mathrm{wait}$, as the time elapsed between the peak of the
previous burst and the peak of a given burst, available when no
data gaps were present before the burst. We measured the
daily-averaged burst recurrence time, $t_\mathrm{rec}$, as the total exposure
time during one day (excluding those orbits where no bursts are
detected) divided by the number of bursts detected on that
day. Therefore, the instantaneous and daily-averaged burst rate,
$\nu_\mathrm{burst}$, are given by $t_\mathrm{wait}^{-1}$ and $t_\mathrm{rec}^{-1}$,
respectively. When only one burst was detected on a given day, we
considered $t_\mathrm{rec}$ an approximate lower limit on the recurrence
time. We also obtained daily averages of $E_\mathrm{b}$, peak burst luminosity
($L_\mathrm{peak}$), blackbody temperature (k$T_\mathrm{peak}$) and radius
($R_\mathrm{peak}$), following the same method described in
\citet{Linares11b}: peak burst values correspond to a 4~s long
interval around the burst peak.

%preprint:
%\begin{table}[b]
%apjlet:
\begin{table*}[t!]
%\rotate
%\scriptsize
%\tiny
\caption{Daily-averaged burst properties and persistent luminosity from T5X2.}
\begin{minipage}{\textwidth}
\begin{center}
\begin{tabular}{ l c c c c c c c c}
\hline\hline
Date\footnote{MJD~55482 is 2010 October 13, and MJD~55519 is 2010 November 19.} & Exposure\footnote{Total daily exposure time in data segments (orbits) where bursts are detected. The total daily exposure time for segments without detected bursts is indicated between square brackets (whenever this time is larger than the burst recurrence time). Thus square brackets indicate periods of intrinsic burst cessation.} & Bursts\footnote{Total number of bursts detected per day. Square brackets indicate periods of intrinsic burst cessation, when no bursts were detected despite long enough exposure time.} & $t_\mathrm{rec}$\footnote{Daily-averaged burst recurrence time.} & $E_\mathrm{b}$\footnote{Bolometric integrated burst energy ($E_\mathrm{b}$) and peak burst luminosity ($L_\mathrm{peak}$). Persistent luminosity ($L_\mathrm{2-50}$) in the 2--50~keV energy band (see Sec.~\ref{sec:data} for bolometric correction). Peak burst blackbody radius ($R_\mathrm{peak}$) not color- or redshift-corrected. All use a distance of 6.3~kpc.} & $L_\mathrm{peak}$$^e$ & k$T_\mathrm{peak}$\footnote{Peak burst blackbody temperature.} & $R_\mathrm{peak}$$^e$ & $L_\mathrm{2-50}$$^e$ \\
(MJD) & (s) &  & (s) & 10$^{39}$erg & 10$^{37}$erg/s & (keV) & (km) & 10$^{37}$erg/s \\
\hline
55482 & 3158 & 1 & 3158.0 & 1.60 $\pm$ 0.07 & 4.60 $\pm$ 0.57 & 2.27 $\pm$ 0.08 & 3.67 $\pm$ 0.23 &  2.20 $\pm$ 0.25 \\
55483 & 20666 & 24 & 861.1 & 0.42 $\pm$ 0.01 & 2.00 $\pm$ 0.10 & 2.13 $\pm$ 0.03 & 2.70 $\pm$ 0.07 &  5.70 $\pm$ 0.28 \\
55484 & 19709 & 42 & 469.3 & 0.28 $\pm$ 0.01 & 1.30 $\pm$ 0.07 & 2.01 $\pm$ 0.03 & 2.35 $\pm$ 0.07 &  6.60 $\pm$ 0.36 \\
55485 & 16514 & 48 & 344.0 & 0.29 $\pm$ 0.01 & 1.30 $\pm$ 0.07 & 1.91 $\pm$ 0.02 & 2.69 $\pm$ 0.07 &  8.30 $\pm$ 0.39 \\
55486 & [17334] & [0] & - & - & - & - & - &  9.55 $\pm$ 0.44 \\
55487 & 14235 [6980] & 50 [0] & 284.7 & 0.36 $\pm$ 0.01 & 1.30 $\pm$ 0.08 & 1.72 $\pm$ 0.02 & 3.01 $\pm$ 0.10 &  10.00 $\pm$ 0.45 \\
55488 & 11755 [3633] & 47 [0] & 250.1 & 0.37 $\pm$ 0.01 & 1.20 $\pm$ 0.07 & 1.69 $\pm$ 0.02 & 3.13 $\pm$ 0.09 &  9.20 $\pm$ 0.46 \\
55489 & 3565 [3752] & 13 [0] & 274.2 & 0.31 $\pm$ 0.01 & 1.00 $\pm$ 0.11 & 1.77 $\pm$ 0.05 & 2.48 $\pm$ 0.14 &  8.40 $\pm$ 0.43 \\
55490 & 7172 & 21 & 341.5 & 0.28 $\pm$ 0.01 & 1.10 $\pm$ 0.09 & 1.82 $\pm$ 0.03 & 2.65 $\pm$ 0.10 &  8.00 $\pm$ 0.38 \\
55491 & 2410 & 6 & 401.7 & 0.23 $\pm$ 0.01 & 0.99 $\pm$ 0.16 & 1.60 $\pm$ 0.06 & 3.21 $\pm$ 0.26 &  7.10 $\pm$ 0.45 \\
55492 & 6079 & 13 & 467.6 & 0.30 $\pm$ 0.01 & 1.30 $\pm$ 0.14 & 1.80 $\pm$ 0.05 & 2.75 $\pm$ 0.16 &  7.10 $\pm$ 0.37 \\
55493 & 5780 & 10 & 578.0 & 0.24 $\pm$ 0.01 & 1.10 $\pm$ 0.13 & 1.59 $\pm$ 0.04 & 3.28 $\pm$ 0.20 &  6.50 $\pm$ 0.37 \\
55494 & 5963 & 12 & 496.9 & 0.27 $\pm$ 0.01 & 1.20 $\pm$ 0.14 & 1.73 $\pm$ 0.05 & 2.95 $\pm$ 0.18 &  6.30 $\pm$ 0.37 \\
55495 & 6685 & 10 & 668.5 & 0.51 $\pm$ 0.01 & 1.70 $\pm$ 0.11 & 1.90 $\pm$ 0.03 & 3.10 $\pm$ 0.11 &  6.20 $\pm$ 0.38 \\
55496 & 6860 & 11 & 623.6 & 0.36 $\pm$ 0.01 & 1.30 $\pm$ 0.11 & 1.83 $\pm$ 0.04 & 2.79 $\pm$ 0.12 &  6.10 $\pm$ 0.36 \\
55497 & 5964 & 8 & 745.5 & 0.37 $\pm$ 0.02 & 1.60 $\pm$ 0.15 & 1.97 $\pm$ 0.04 & 2.76 $\pm$ 0.13 &  5.90 $\pm$ 0.38 \\
55498 & 5150 & 6 & 858.3 & 0.48 $\pm$ 0.02 & 1.60 $\pm$ 0.14 & 1.94 $\pm$ 0.04 & 2.95 $\pm$ 0.13 &  5.60 $\pm$ 0.42 \\
55499 & 5364 & 7 & 766.3 & 0.35 $\pm$ 0.01 & 1.60 $\pm$ 0.13 & 1.96 $\pm$ 0.04 & 2.81 $\pm$ 0.11 &  5.50 $\pm$ 0.34 \\
55500 & 5372 & 4 & 1343.0 & 0.74 $\pm$ 0.03 & 1.90 $\pm$ 0.17 & 1.98 $\pm$ 0.04 & 3.11 $\pm$ 0.14 &  5.50 $\pm$ 0.32 \\
55501 & 5236 & 5 & 1047.2 & 0.79 $\pm$ 0.02 & 2.10 $\pm$ 0.16 & 2.02 $\pm$ 0.04 & 3.07 $\pm$ 0.11 &  5.50 $\pm$ 0.33 \\
55502 & 6789 & 6 & 1131.5 & 0.87 $\pm$ 0.02 & 2.50 $\pm$ 0.15 & 2.06 $\pm$ 0.03 & 3.19 $\pm$ 0.10 &  5.20 $\pm$ 0.30 \\
55503 & 5292 & 5 & 1058.4 & 0.74 $\pm$ 0.02 & 2.30 $\pm$ 0.16 & 2.11 $\pm$ 0.04 & 3.00 $\pm$ 0.10 &  5.30 $\pm$ 0.37 \\
55504 & 6810 & 6 & 1135.0 & 0.57 $\pm$ 0.02 & 2.10 $\pm$ 0.14 & 2.15 $\pm$ 0.04 & 2.75 $\pm$ 0.09 &  5.20 $\pm$ 0.31 \\
55505 & 5536 & 4 & 1384.0 & 0.74 $\pm$ 0.02 & 2.20 $\pm$ 0.18 & 2.12 $\pm$ 0.04 & 2.88 $\pm$ 0.12 &  5.20 $\pm$ 0.32 \\
55506 & 6803 & 5 & 1360.6 & 0.93 $\pm$ 0.02 & 2.60 $\pm$ 0.17 & 2.14 $\pm$ 0.04 & 3.06 $\pm$ 0.10 &  5.10 $\pm$ 0.31 \\
55508 & 6799 & 4 & 1699.8 & 1.00 $\pm$ 0.03 & 2.70 $\pm$ 0.19 & 2.18 $\pm$ 0.04 & 3.07 $\pm$ 0.10 &  4.90 $\pm$ 0.30 \\
55509 & 6781 & 5 & 1356.2 & 1.06 $\pm$ 0.02 & 2.90 $\pm$ 0.16 & 2.21 $\pm$ 0.03 & 3.08 $\pm$ 0.08 &  4.70 $\pm$ 0.29 \\
55510 & 4893 & 3 & 1631.0 & 1.02 $\pm$ 0.03 & 3.10 $\pm$ 0.23 & 2.08 $\pm$ 0.04 & 3.54 $\pm$ 0.13 &  4.70 $\pm$ 0.31 \\
55511 & 6788 & 4 & 1697.0 & 1.26 $\pm$ 0.03 & 3.10 $\pm$ 0.20 & 2.07 $\pm$ 0.03 & 3.60 $\pm$ 0.11 &  4.60 $\pm$ 0.29 \\
55512 & 5329 & 3 & 1776.3 & 1.28 $\pm$ 0.03 & 3.30 $\pm$ 0.21 & 2.17 $\pm$ 0.04 & 3.35 $\pm$ 0.11 &  4.60 $\pm$ 0.31 \\
55513 & 5661 & 2 & 2830.5 & 1.37 $\pm$ 0.05 & 3.30 $\pm$ 0.37 & 2.30 $\pm$ 0.07 & 3.01 $\pm$ 0.17 &  4.60 $\pm$ 0.34 \\
55514 & 6737 & 2 & 3368.5 & 1.46 $\pm$ 0.03 & 3.70 $\pm$ 0.26 & 2.25 $\pm$ 0.04 & 3.32 $\pm$ 0.12 &  4.40 $\pm$ 0.28 \\
55515 & 10197 & 4 & 2549.2 & 1.33 $\pm$ 0.02 & 3.70 $\pm$ 0.18 & 2.23 $\pm$ 0.03 & 3.40 $\pm$ 0.08 &  4.30 $\pm$ 0.29 \\
55516 & 6803 & 4 & 1700.8 & 1.37 $\pm$ 0.03 & 3.70 $\pm$ 0.25 & 2.16 $\pm$ 0.04 & 3.61 $\pm$ 0.12 &  4.20 $\pm$ 0.24 \\
55517 & 3410 & 1 & 3410.0 & 1.42 $\pm$ 0.04 & 3.70 $\pm$ 0.29 & 2.21 $\pm$ 0.04 & 3.47 $\pm$ 0.13 &  4.00 $\pm$ 0.28 \\
55518 & 3407 & 2 & 1703.5 & 1.22 $\pm$ 0.03 & 3.80 $\pm$ 0.26 & 2.31 $\pm$ 0.04 & 3.24 $\pm$ 0.11 &  4.00 $\pm$ 0.30 \\
55519 & 6788 & 2 & 3394.0 & 1.45 $\pm$ 0.04 & 3.90 $\pm$ 0.34 & 2.19 $\pm$ 0.05 & 3.61 $\pm$ 0.16 &  3.80 $\pm$ 0.27 \\
\hline\hline
\end{tabular}
\end{center}
\end{minipage}
\label{table:dailybursts}
%preprint:
%\end{table}
%apjlet:
\end{table*}

Due to the smooth evolution from a series of bursts into a mHz QPO and
vice versa (see Fig.~\ref{fig:evol} and Sec.~\ref{sec:results}) the
distinction between ``frequent bursts'' and mHz QPOs is, to some
extent, arbitrary. We searched for mHz QPOs all observations taken
when the daily-averaged burst recurrence time was shorter than 350~s,
which corresponds to 10 observations between MJDs~55485--55490 (around
the outburst peak). Given the typical duration of a continuous {\it
RXTE} observation segment (an ``orbit''), this threshold ensures that
about 10 or more QPO cycles, or bursts, are observed without
interruption. To study the mHz QPOs in those {\it RXTE} orbits we used
2--60 keV 1~s-bin light curves from all active PCUs combined. We then
calculated a Lomb-Scargle periodogram \citep[LSP, oversampled by a
factor of 3; ][]{Lomb76,Scargle82} for each light curve and measured
the mHz QPO frequency, $\nu_\mathrm{QPO}$, as that frequency with the
highest power in the periodogram. The corresponding period was then
used to fold the background-corrected light curve (using background
rates from {\it pcabackest} in 16-s steps) and produce a mHz QPO
folded profile, from which we measured the fractional
root-mean-squared (rms) amplitude. To investigate the energy
dependence of the mHz QPO amplitude, we also produced light curves in
five different energy bands (in keV: 2--3, 3--5.5, 5.5--9.5, 9.5--21
and 21--53), folded the light curve at the period found in the
respective 2--60 keV range dataset and measured the ``rms spectrum''
of the mHz QPOs. We also performed 2048~s-long fast Fourier transforms
(FFTs) using the 2--60~keV energy band and the same {\it RXTE} orbits,
in order to constrain the mHz QPO coherence or ``quality factor''.

\begin{figure}[h]
\centering
%  \begin{center}
  \resizebox{1.0\columnwidth}{!}{\rotatebox{0}{\includegraphics[]{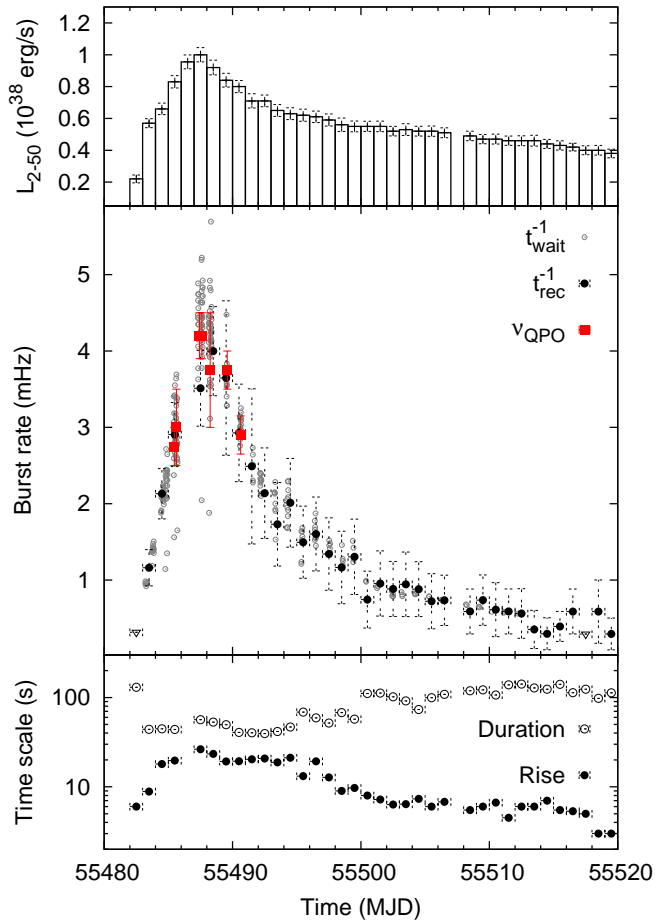}}}
  \caption{From top to bottom panels, evolution along the outburst of
  {\it i)} persistent 2--50~keV luminosity; {\it ii)} burst rate as
  measured from $t_\mathrm{wait}$ (open gray circles), $t_\mathrm{rec}$ (filled
  black circles) and mHz QPO frequency (red filled squares; see
  Sec.~\ref{sec:data} for definitions) and {\it iii)} burst rise time
  and duration. Gray and black circles show individual burst
  measurements and daily averages, respectively. Open triangles show
  burst rate daily averages based on one single burst, which we
  consider as upper limits.}
    \label{fig:ob}
% \end{center}
\end{figure}

In order to measure the persistent (accretion) luminosity, we
extracted one dead-time-corrected spectrum per observation from
Standard~2 data, excluding all bursts and subtracting the background
spectrum estimated with the bright source background model and {\it
pcabackest} (v. 3.8). We then fitted each persistent spectrum with a
model consisting of a disk blackbody, a power law and a Gaussian line
with energy fixed at 6.5~keV, correcting for absorption as above. We
calculated the 2--50~keV persistent luminosity ($L_\mathrm{2-50}$)
from the best fit model. Furthermore, we measured the 0.01--2~keV
unabsorbed flux from a simultaneous fit to the T5X2 spectrum measured
by {\it Swift}-XRT (0.5-10~keV) and {\it RXTE}-PCA (2.5-25~keV) on MJD
55501 (2010 November 01), extrapolating the {\it phabs*simpl(bbody +
diskbb)} best fit model down to 0.01~keV, and found a bolometric
correction factor of 1.13. This bolometric correction factor converts
2--50~keV into 0.01--50~keV flux, which we take as bolometric flux
given that the persistent spectrum remains soft (photon index
2.4--3.3) throughout the outburst and we therefore do not expect
sizeable emission above 50~keV (with the only exception of the first
observation, when T5X2 was in the hard state and the photon index was
$\sim$1.7). Using the same procedure, we found a bolometric correction
factor of 1.02 for the 0.5--50~keV luminosity from Cir~X-1 reported by
\citet[][used in Sec.~\ref{sec:discussion}]{Linares10d}. We applied
these bolometric correction factors in order to estimate the
bolometric persistent luminosity: $L_\mathrm{pers}$. We use throughout
this work an Eddington luminosity of $L_\mathrm{Edd}$=$2.5\times
10^{38}$ erg s$^{-1}$ to calculate Eddington-normalized
$L_\mathrm{pers}$. To convert from $L_{pers}$ to $\dot{m}$, we assume
homogeneous accretion onto the NS and that the
(general-relativistic) gravitational energy of the in-falling
matter is fully converted into radiation at the surface of a
1.4~$M_\odot$ mass, 10~km radius NS. We define the Eddington mass
accretion rate per unit area, $\dot{m}_\mathrm{Edd}$, as the mass
accretion rate needed to sustain a luminosity equal to
$L_\mathrm{Edd}$. With this definition
$\dot{m}_\mathrm{Edd}$=$1.2\times 10^5$~g~cm$^{-2}$~s$^{-1}$.

 \begin{table}[b]
%\scriptsize
\tiny
\caption{Properties of the mHz QPOs from T5X2.}
\footnotetext{For comparison, the reported values for the
``low-$L_\mathrm{pers}$ mHz QPOs'' are in the following ranges:
$\nu_\mathrm{QPO}$=[7--14.3]~mHz; rms$_\mathrm{QPO}$=[0.7--1.9]\%
(2--5~keV); $L_\mathrm{pers}$=[0.6--3.5]$\times$10$^{37}$erg/s
\citep[Sec.~\ref{sec:msb};][]{Revnivtsev01,Altamirano08d}.}
\footnotetext{\footnotemark{}Observation (from proposal-target 95437-01) where the
mHz QPOs are detected, orbit numbers used indicated between square
brackets.}
\begin{minipage}{\textwidth}
\begin{tabular}{ l l c c c c}
\hline\hline
Date & OBSID\footnotemark{} & $\nu_\mathrm{QPO}$ & rms$_\mathrm{QPO}$ & $L_\mathrm{pers}$/10$^{37}$ & $\dot{m}$/10$^{4}$  \\
(MJD) &  & (mHz) & (\%) & (erg/s) &
(g~cm$^{-2}$s$^{-1}$) \\
\hline
55485.46 & 04-00 [1]  & 2.75 $\pm$ 0.2 & 1.9 $\pm$ 0.1 & 9.1 $\pm$   0.4  & 4.5 $\pm$ 0.2 \\
55485.63 & 04-01 [1,2,3]  & 3 $\pm$ 0.5     & 2.2 $\pm$ 0.1 & 9.9 $\pm$  0.4  & 4.9 $\pm$ 0.2 \\
55487.43 & 06-000 [1,4] & 4.2 $\pm$ 0.2   & 1.3 $\pm$ 0.1 & 11.4 $\pm$ 0.4 & 5.6 $\pm$ 0.2 \\
55487.62 & 06-00 [1,2] & 4.2 $\pm$ 0.2   & 1.7 $\pm$ 0.2 & 12.1 $\pm$   0.5 & 6.0 $\pm$ 0.2 \\
55488.26 & 07-00 [1] & 4.0 $\pm$ 0.5   & 1.4 $\pm$ 0.1 & 12.0 $\pm$ 0.5 & 5.9 $\pm$ 0.2 \\
55489.59 & 08-00 [1] & 3.75 $\pm$ 0.2 & 2.1 $\pm$ 0.1 & 9.6 $\pm$ 0.5 & 4.8 $\pm$ 0.2 \\
55490.63 & 09-00 [1,2] & 2.9 $\pm$ 0.2  & 2.2 $\pm$ 0.1 & 9.1 $\pm$ 0.4  & 4.5 $\pm$ 0.2 \\
\hline\hline
\end{tabular}
\end{minipage}
\label{table:mhzqpo}
\end{table}

\section{Results}
\label{sec:results}

We found and analyzed a total of 398 X-ray bursts occurring between
2010 October 13 and 2010 November 19, including the faint and frequent bursts
near the peak of the outburst that form the mHz QPOs (see below). We
show in Figure~\ref{fig:evol} an overview of the joint evolution of
persistent emission and burst properties along the T5X2
outburst. Bursts become more frequent and fainter when the persistent
luminosity, $L_\mathrm{pers}$, increases, and they turn into brighter
and more frequent bursts during the outburst decay. Most
interestingly, the X-ray bursts gradually develop into the observed
mHz QPO during the outburst rise, and the mHz QPO mutates into a
series of bursts along the outburst decay (Figure~\ref{fig:evol}), an
unprecedented phenomenon among thermonuclear bursters. This also
makes T5X2 the most prolific source of thermonuclear bursts known to
date (an average burst rate over more than a month of 4.9 hr$^{-1}$;
5.5 hr$^{-1}$ excluding orbits where no bursts were detected), with
the shortest recurrence times between thermonuclear bursts observed to
date \citep[as short as $\sim$200~s; c.f.][]{Linares09c,Keek10}.

The {\it smooth metamorphosis} between bursts and mHz QPOs can be seen
qualitatively in Figure~\ref{fig:evol}, and quantitatively by studying
the evolution of several burst properties (Figure~\ref{fig:ob} \&
Table~\ref{table:dailybursts}). Burst rate, rise time and duration as
well as peak burst luminosity, $L_\mathrm{peak}$, and total radiated
energy $E_\mathrm{b}$; \citep[see also][]{Linares11b}, all evolve
gradually while $L_\mathrm{pers}$ changes by a factor of $\sim$5 along
the outburst. Figure~\ref{fig:ob} (middle panel) also shows that the
burst rate equals the mHz QPO frequency during the outburst peak,
$\nu_\mathrm{burst}$=$\nu_\mathrm{QPO}$, as expected given that the
mHz QPOs are simply formed by a series of faint and frequent
bursts. This unique behavior makes the distinction between bursts and
mHz QPOs somewhat arbitrary. As explained in Section~\ref{sec:data},
our practical definition of mHz QPO requires burst recurrence times
shorter than 350~s, so that typically 10 or more QPO cycles are
observed without interruption. In a few occasions we find that
one burst is missing from the series of regular bursts, and the
corresponding values of $\nu_\mathrm{burst} = t_\mathrm{wait}^{-1}$ for
individual bursts are a factor $\sim$2 lower than the general trend,
as can be seen in Figure~\ref{fig:ob} (middle panel). Interestingly, a
similar behavior with sporadic ``missing bursts'' is seen in the mHz
QPO simulations presented in \citet{Heger07}.

\begin{figure}[h!]
  \begin{center}
  \resizebox{1.0\columnwidth}{!}{\rotatebox{-90}{\includegraphics[]{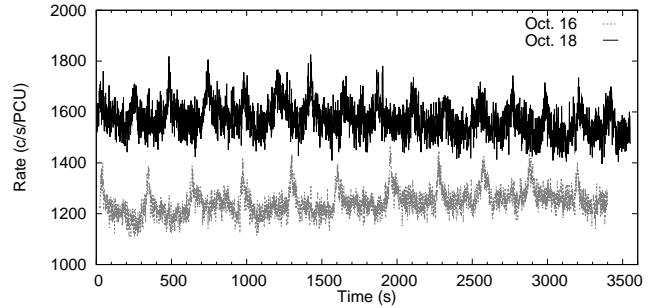}}}
  \caption{Light curves of the mHz QPOs on October 16 (gray) and 18
  (black), in the 2--60~keV energy band, using 1-s time bins and no
  background subtraction. One {\it RXTE} orbit is shown in each case,
  and time is relative to the start of that orbit (2010-10-16 14:33:27
  UTC and 2010-10-18 15:09:55 UTC). The persistent luminosity was
  higher on October 18 ($L_\mathrm{pers}$$\simeq$0.45$L_\mathrm{Edd}$)
  than October 16 ($L_\mathrm{pers}$$\simeq$0.38$L_\mathrm{Edd}$). To
  illustrate this, count rate is normalized to the number of active
  PCUs.}
    \label{fig:qpoLC}
 \end{center}
\end{figure}

\begin{figure}[h!]
  \begin{center}
  \resizebox{1.0\columnwidth}{!}{\rotatebox{-90}{\includegraphics[]{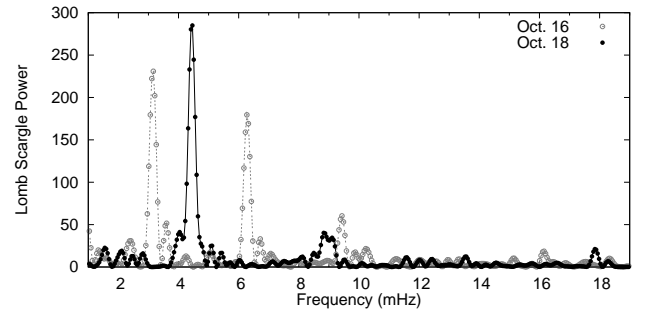}}}
  \caption{Lomb-Scargle periodograms of the mHz QPOs on October 16
  (gray) and 18 (black), showing the QPO harmonic structure and the
  change in $\nu_\mathrm{QPO}$.}
    \label{fig:qpoLS}
 \end{center}
\end{figure}

\begin{figure}[h!]
  \begin{center}
  \resizebox{1.0\columnwidth}{!}{\rotatebox{-90}{\includegraphics[]{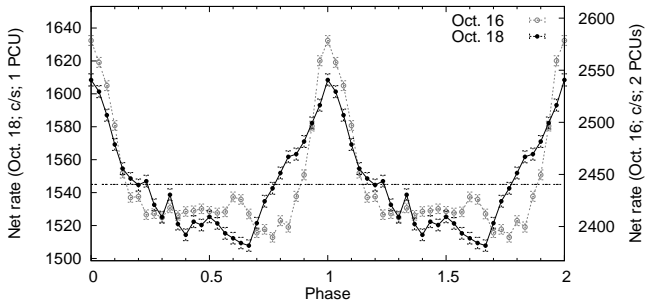}}}
  \caption{Folded background-subtracted 2--60~keV light curves of the
  mHz QPOs on October 16 (gray) and 18 (black). Peaks are at phase 0
  by definition. Two QPO cycles are shown for clarity.  Two PCUs were
  active on October 16, yielding a higher collected rate. The dashed
  horizontal line shows the average count rate, and the vertical range
  corresponds to 10\% of that value in each case, showing that the
  fractional amplitude was higher on October 16
  (Table~\ref{table:mhzqpo}). The net peak burst luminosity was in
  both cases $\simeq$1.3$\times$10$^{37}$ erg~s$^{-1}$
  (Table~\ref{table:dailybursts}).}
    \label{fig:qpoFP}
%\epsscale{1.0}
 \end{center}
\end{figure}

\begin{figure}[h!]
  \begin{center}
  \resizebox{1.0\columnwidth}{!}{\rotatebox{-90}{\includegraphics[]{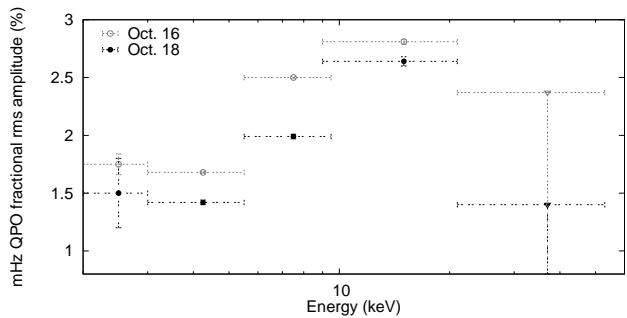}}}
  \caption{``RMS spectrum'' of the mHz QPOs on October 16 (gray) and
  18 (black). Their fractional rms amplitude increases with energy up
  to $\sim$20~keV. No mHz QPOs are detected above that energy, empty
  triangles show upper limits on their fractional rms amplitude.}
    \label{fig:qpoRE}
 \end{center}
\end{figure}

\subsection{mHz QPOs}
\label{sec:mhzqpo}

We report the discovery of several instances of mHz QPOs during the
peak of the T5X2 outburst, on 2010 October 16, 18, 19, 20 and 21
\citep[MJDs~55485--55490;][for the initial report of mHz QPOs on 2010
October 18 and 19]{Linares10c}. Two examples of mHz QPO light curves
are shown in Figure~\ref{fig:qpoLC}, each spanning one {\it RXTE}
orbit. Table~\ref{table:mhzqpo} shows the main QPO properties:
fractional rms amplitudes between 1.3\% and 2.2\% (in the 2--60~keV
band) and $\nu_\mathrm{QPO}$ in the range 2.8--4.2~mHz (note that the
lower end of this frequency range corresponds to our mHz QPO
definition, Secs.~\ref{sec:data} \&
\ref{sec:results}). Figure~\ref{fig:qpoLS} presents LSPs of two cases,
on October 16 and 18, clearly showing the change in $\nu_\mathrm{QPO}$
as well as the harmonic structure (up to 4 overtones are visible in
the LSPs). By inspecting the power spectra obtained from 2048-s-long
FFTs, we find that the mHz QPO power is in all cases spread over one
or two frequency bins, from which we derive a limit on the
full-width-at-half-maximum, FWHM$\lesssim$1~mHz. For the 2.8--4.2~mHz
QPO frequencies this corresponds to a lower limit on the quality
factor (Q$\equiv$$\nu_\mathrm{QPO}$/FWHM) of Q$\gtrsim$3, which
reveals a fairly coherent QPO (a fairly constant $t_\mathrm{wait}$).

We present in Figure~\ref{fig:qpoFP} light curves folded at the mHz
QPO period, showing folded burst/QPO profiles for the same two dates,
October 16 and 18. In both cases the burst/QPO profile is peaked and
highly non-sinusoidal (as seen also in the raw, unfolded light curves;
Fig.~\ref{fig:qpoLC}), which explains the harmonic content. On October
18, when $\nu_\mathrm{QPO}$ was highest (Table~\ref{table:mhzqpo}), we
find a nearly symmetric burst/QPO profile. On October 16 the burst/QPO
profile is slightly asymmetric, with the rise faster than the decay.
We measure a mHz QPO fractional rms amplitude in the range 1.3--2.2
\%, in the total ($\sim$2--60~keV) PCA band. Furthermore, from our
measurements of the mHz QPO amplitude at different energies
(Sec.~\ref{sec:data}; Fig.~\ref{fig:qpoRE}) we find that its
fractional rms amplitude increases between $\sim$2 and $\sim$20~keV,
from $\sim$1.4\% to $\sim$2.8\% in the two cases presented in
Figure~\ref{fig:qpoRE}. At energies higher than 20~keV the mHz QPO is
not visible by naked eye in the light curves, nor it is conclusively
detected using LSPs, FFTs or folded light curves. The presence of red
noise with variable strength gives rise to different upper limits on
the fractional rms amplitude of the mHz QPO above 20~keV (between
1.4\% and 3\%; Fig.~\ref{fig:qpoRE}). As discussed in
Section~\ref{sec:msb}, such increase in QPO amplitude up to 20~keV is
in contrast with mHz QPOs from other bursting NSs, which showed a
fractional amplitude decreasing with increasing energy between 2 and 5
keV \citep{Revnivtsev01}.

\begin{figure}[h!]
  \begin{center}
  \resizebox{1.0\columnwidth}{!}{\rotatebox{-90}{\includegraphics[]{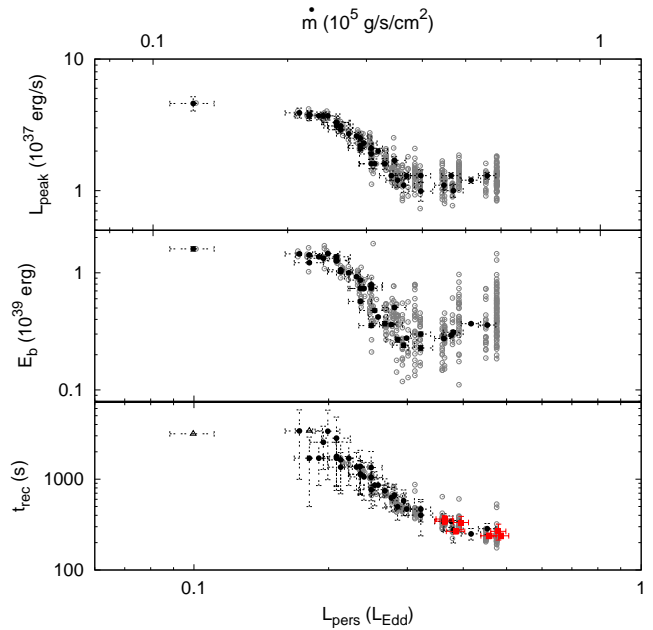}}}
  \caption{Burst peak luminosity {\it (top)}, integrated burst energy
  {\it (middle)} and burst recurrence time, $t_\mathrm{rec}$ {\it
  (bottom)} vs.  bolometric persistent luminosity ($L_\mathrm{pers}$).
  Gray and black circles show individual burst measurements and daily
  averages, respectively. Red filled squares show recurrence times
  measured from the inverse of the mHz QPO frequency. The top axis
  shows the inferred $\dot{m}$ assuming homogeneous accretion onto a
  1.4~$M_\odot$, 10~km radius NS (Sec.~\ref{sec:data}).}
    \label{fig:lumlum}
 \end{center}
\end{figure}

\begin{figure}[h!]
  \begin{center}
  \resizebox{1.0\columnwidth}{!}{\rotatebox{-90}{\includegraphics[]{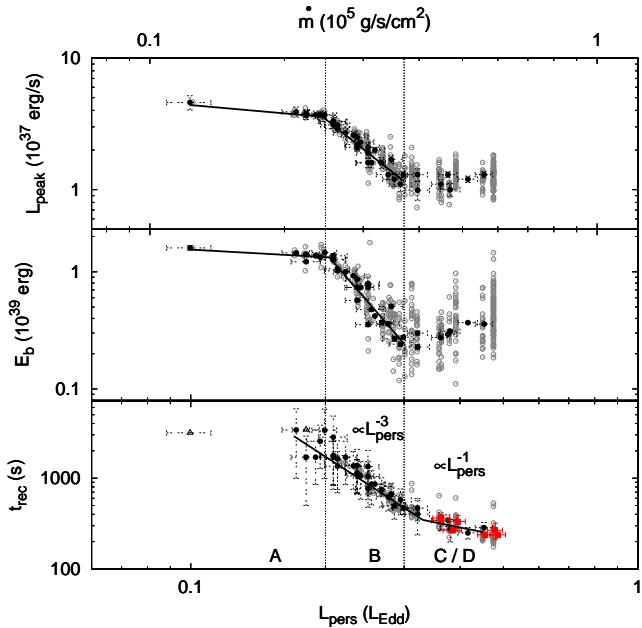}}}
  \caption{Same as Figure~\ref{fig:lumlum}, showing also the broken
power law fits to the
$L_\mathrm{peak}$, $E_\mathrm{b}$, $t_\mathrm{rec}$
vs. $L_\mathrm{pers}$ relations (Table~\ref{table:power}). Three
regimes are apparent and labeled along the bottom axis: slow decrease
(at $L_\mathrm{pers}$/$L_\mathrm{Edd}$$\lesssim$0.2; regime A), fast
drop (at 0.2$\lesssim$$L_\mathrm{pers}$/$L_\mathrm{Edd}$$\lesssim$0.3;
regime B) and saturation
($L_\mathrm{pers}$/$L_\mathrm{Edd}$$\gtrsim$0.3; regime C) of the
$L_\mathrm{peak}$, $E_\mathrm{b}$ vs. $L_\mathrm{pers}$ relations. The
$t_\mathrm{rec}$ values follow approximate
$t_\mathrm{rec}$~$\propto$~$L_\mathrm{pers}^{-3}$ and
$t_\mathrm{rec}$~$\propto$~$L_\mathrm{pers}^{-1}$ relations in regimes
B and C, as indicated (see Figure~\ref{fig:trec},
Table~\ref{table:regimes} and Secs.~\ref{sec:burstmdot} \&
\ref{sec:discussion}).}
    \label{fig:lumlumlin}
%\epsscale{1.0}
 \end{center}
\end{figure}

\begin{figure}[h!]
  \begin{center}
  \resizebox{1.0\columnwidth}{!}{\rotatebox{-90}{\includegraphics[]{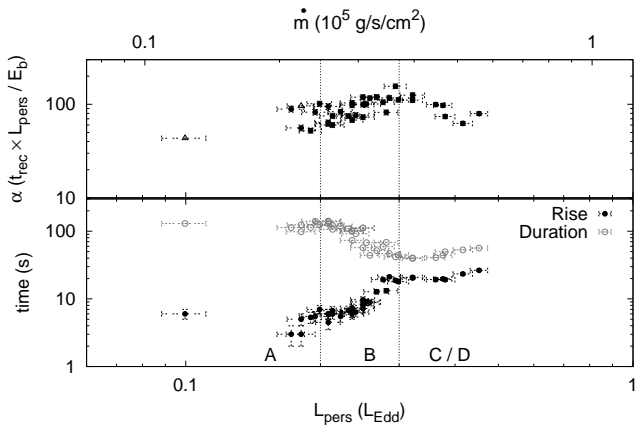}}}
  \caption{Daily-averaged accretion-to-burst energy ratio ($\alpha$,
  top panel) and burst time scales (rise time and duration, bottom
  panel) vs. persistent luminosity ($L_\mathrm{pers}$) during the T5X2
  outburst. Open triangles represent lower limits (one single burst
  detected during that day). The corresponding bursting regimes
  are indicated with dotted lines and labels along the bottom axis
  (Table~\ref{table:regimes}).}
    \label{fig:alpha}
 \end{center}
\end{figure}

\begin{table}[ht]
%\scriptsize
%\tiny
\caption{Broken power law fits to the burst properties vs. persistent luminosity relations.}
\begin{minipage}{\textwidth}
\begin{tabular}{l c c c}
\hline\hline
Relation\footnote{Pairs of variables fitted with a broken power law: $L_\mathrm{peak}$, $E_\mathrm{b}$\\ and $t_\mathrm{rec}$ vs. $L_\mathrm{pers}$ (see Figures~\ref{fig:lumlum} \& \ref{fig:lumlumlin}). The $L_\mathrm{pers}$ range used in\\ each fit is quoted between brackets, in units of $L_\mathrm{Edd}$. i$_1$ and i$_2$\\ represent, respectively, the power law indices before and after the\\ break (at $L_\mathrm{break}$).}  &$L_\mathrm{peak}$-$L_\mathrm{pers}$ &$E_\mathrm{b}$-$L_\mathrm{pers}$ &$t_\mathrm{rec}$-$L_\mathrm{pers}$ \\ 
$[$range$]$ & [0.1--0.3] & [0.1--0.3] & [0.17--0.45] \\
\hline
 K\footnote{Power law normalization before the break, same units as $L_\mathrm{peak}$\\ (10$^{37}$ erg~s$^{-1}$), $E_\mathrm{b}$ (10$^{39}$ erg) and $t_\mathrm{rec}$ (s).} & 2.2 $\pm$ 0.7 & 0.9 $\pm$ 0.1 & 10 $\pm$ 7 \\ 
 i$_1$ & -0.3 $\pm$ 0.2 & -0.2 $\pm$ 0.1 & -3.2 $\pm$ 0.5 \\ 
 $L_\mathrm{break}$($L_\mathrm{Edd}$) & 0.20 $\pm$ 0.01 & 0.20 $\pm$ 0.01 & 0.33 $\pm$ 0.04 \\ 
 i$_2$ & -2.7 $\pm$ 0.1 & -4.4 $\pm$ 0.1 & -1.0 $\pm$ 0.9 \\ 
 $\chi^2$ / dof& 38.5 / 25 & 930 / 25 & 6.1 / 30 \\ 
\hline\hline
\end{tabular}
\end{minipage}
\label{table:power}
\end{table}

\subsection{Burst properties vs. accretion rate}
\label{sec:burstmdot}

We present in this Section the relation between burst properties and
$\dot{m}$, as inferred from $L_\mathrm{pers}$, including the faintest
and most frequent bursts which form the mHz QPOs (see above).
Figure~\ref{fig:ob} shows $L_\mathrm{peak}$, $E_\mathrm{b}$ and
$t_\mathrm{rec}$ as a function of the Eddington-normalized
$L_\mathrm{pers}$ and $\dot{m}$ (Sec.~\ref{sec:data}). The overall
trend is that of an anti-correlation between $L_\mathrm{peak}$,
$E_\mathrm{b}$ and $t_\mathrm{rec}$ on the one hand and
$L_\mathrm{pers}$ on the other (see also Fig.~1 in
\citealt{Linares11b}; \citealt{Motta11}). Closer inspection of the
burst properties over the full $L_\mathrm{pers}$ range
(0.1--0.5~$L_\mathrm{Edd}$; see Figure~\ref{fig:lumlum}) reveals a
more complex and interesting behavior, namely, four different bursting
regimes.

\begin{itemize}

\item At the lowest persistent luminosities,
  0.1$<$$L_\mathrm{pers}$/$L_\mathrm{Edd}$$<$0.2, when
  $L_\mathrm{pers}$ increases both $L_\mathrm{peak}$ and
  $E_\mathrm{b}$ (and possibly $t_\mathrm{rec}$) decrease moderately,
  from $L_\mathrm{peak}$$\simeq$4.6$\times 10^{37}$erg/s,
  $E_\mathrm{b}$$\simeq$1.6$\times 10^{39}$ erg to
  $L_\mathrm{peak}$$\simeq$3.7$\times 10^{37}$erg/s,
  $E_\mathrm{b}$$\simeq$1.3$\times 10^{39}$erg. We refer to this
  0.1--0.2~$L_\mathrm{Edd}$ regime as {\it slow decrease, or regime A}
  (see Fig.~\ref{fig:lumlum}).

\item At higher persistent luminosities,
  0.2$<$$L_\mathrm{pers}$/$L_\mathrm{Edd}$$<$0.3, we find
  $L_\mathrm{peak}$ and $E_\mathrm{b}$ to be steeply anticorrelated
  with $L_\mathrm{pers}$, while $t_\mathrm{rec}$ also decreases from
  200~s to 400~s with increasing $L_\mathrm{pers}$. We refer to this
  0.2--0.3~$L_\mathrm{Edd}$ regime as {\it fast drop, or regime B}
  (see Fig.~\ref{fig:lumlum}).

\item At the highest persistent luminosities
($L_\mathrm{pers}$$>$0.3~$L_\mathrm{Edd}$), when mHz QPOs are
detected, burst peak luminosity and energy reach approximately
constant values ({\it saturation, or regime C}):
$L_\mathrm{peak}$$\simeq$1.2$\times 10^{37}$erg/s and
$E_\mathrm{b}$$\simeq$0.3$\times 10^{39}$erg. The burst recurrence
time, however, keeps decreasing with increasing $L_\mathrm{pers}$ from
$t_\mathrm{rec}$$\simeq$400~s down to $\sim$240~s on 2010 October 18
(Fig.~\ref{fig:lumlum}; Table~\ref{table:mhzqpo}).

\item Finally, during incursions into the flaring/normal branches near
the outburst peak, when T5X2 showed Z source behavior
\citep{Altamirano10b}, neither bursts nor mHz QPOs were
detected. These few episodes without bursting activity, which we refer
to as {\it regime D}, were associated with moderate ($\sim$30\%) short
(hour-long) {\it drops} in $L_\mathrm{pers}$ and occurred between MJDs
55486 and 55496 (while
$L_\mathrm{pers}$$\gtrsim$0.3~$L_\mathrm{Edd}$). They are noted with
square brackets in Table~\ref{table:dailybursts}.

\end{itemize}

\begin{table*}[ht]
%\centering
%\scriptsize
%\small
\caption{Summary of bursting regimes in T5X2. See Section~\ref{sec:burstmdot} for details.}
\begin{minipage}{\textwidth}
\begin{center}
\begin{tabular}{c c c c c c}
\hline\hline
Regime & $L_\mathrm{pers}$ & $t_\mathrm{rec}$ & $E_\mathrm{b}$ & $L_\mathrm{peak}$ &  Description \\
 & ($L_\mathrm{Edd}$) & (s) & (10$^{39}$ erg) &  (10$^{37}$ erg~s$^{-1}$) &  \\
\hline
A & 0.1--0.2 & $>$2000 & 1.6--1.3 & 4.6--3.7 & \specialcell[t]{Slow decrease of $E_\mathrm{b}$ \& $L_\mathrm{peak}$\\ (and $t_\mathrm{rec}$?) with increasing $L_\mathrm{pers}$.}  \\
 &  &  &  &  &  \\
B & 0.2--0.3 & 2000--400 & 1.3--0.3 & 3.7--1.2 & \specialcell[t]{Fast drop of $t_\mathrm{rec}$, $E_\mathrm{b}$ \& $L_\mathrm{peak}$\\ with increasing $L_\mathrm{pers}$. $t_\mathrm{rec}$~$\propto$~$L_\mathrm{pers}$$^{-3}$.} \\
 &  &  &  &  &  \\
C & 0.3--0.5 & 400--200 & $\sim$0.3 & $\sim$1.2 & \specialcell[t]{Bursts evolve into mHz QPOs with\\ increasing frequency ($t_\mathrm{rec}$~$\propto$~$L_\mathrm{pers}$$^{-1}$).\\ $E_\mathrm{b}$ \& $L_\mathrm{peak}$ ``saturate''.} \\
 &  &  &  &  &  \\
D & 0.3--0.5 & - & - & - & \specialcell[t]{No bursts/mHz QPOs detected on \\ normal \& flaring branch in Z source phase.} \\
\hline\hline
\end{tabular}
\end{center}
\end{minipage}
\label{table:regimes}
\end{table*}

Interestingly, regimes C and D show similar $L_\mathrm{pers}$
(0.3--0.5~$L_\mathrm{Edd}$), but their variability and spectral
properties differ (i.e., they constitute different ``accretion
states''; see Altamirano et al. 2012, in prep.). We summarize the main
bursting properties of all four regimes in Table~\ref{table:regimes}.

In order to quantify the anticorrelations explained above and to
characterize the bursting regimes present in T5X2, we fit the
$L_\mathrm{peak}$, $E_\mathrm{b}$, $t_\mathrm{rec}$
vs. $L_\mathrm{pers}$ relations with broken power law functions. The
results are shown in Figure~\ref{fig:lumlumlin} and
Table~\ref{table:power}, and confirm that the two ``breaks'' or
transitions occur at $L_\mathrm{pers}$$\simeq$0.2~$L_\mathrm{Edd}$ and
$L_\mathrm{pers}$$\simeq$0.3~$L_\mathrm{Edd}$. Due to the large
scatter some of the fits are statistically poor, yet they allow us to
constrain the slope and transition luminosity of the three bursting
regimes (Fig.~\ref{fig:lumlumlin}).
Remarkably, the $t_\mathrm{rec}$-$L_\mathrm{pers}$ relation that we
find in regime B ($L_\mathrm{pers}$$\sim$0.2--0.3~$L_\mathrm{Edd}$) is
far from linear, as would be expected from a
$t_\mathrm{rec}$~$\propto$~$\dot{m}^{-1}$ relation if
$\dot{m}$~$\propto$~$L_\mathrm{pers}$ (See Sec.~\ref{sec:trec} for
further discussion). Instead, it is close to
$t_\mathrm{rec}$~$\propto$~$L_\mathrm{pers}^{-3}$: we measure a
$t_\mathrm{rec}$-$L_\mathrm{pers}$ power law index in regime B of
-3.2$\pm$0.5 (see further discussion in Sec.~\ref{sec:discussion}). We
also calculate the mean alpha parameter
($\alpha$$\equiv$$L_\mathrm{pers}$$\times$$t_\mathrm{rec}$ /
$E_\mathrm{b}$) from the daily-averaged values reported by
\citet{Linares11b} and, after applying the bolometric correction we
obtained from broadband X-ray spectral fits (Sec.~\ref{sec:data}), we
find a mean $\alpha$ of 102, with a standard deviation of
27. Figure~\ref{fig:alpha} shows that $\alpha$ increases from $\sim$60
to $\sim$120 with increasing $L_\mathrm{pers}$ until
$L_\mathrm{pers}$$\simeq$0.3~$L_\mathrm{Edd}$ (regimes A \& B), and
decreases at higher luminosities (regime C) to reach again values
close to 60. Figure~\ref{fig:alpha} also shows the burst duration
and rise time as a function of $L_\mathrm{pers}$: the general trend is
for bursts to become shorter and have slower rise when
$L_\mathrm{pers}$ increases.

\begin{figure}[h!]
  \begin{center}
  \resizebox{1.05\columnwidth}{!}{\rotatebox{-90}{\includegraphics[]{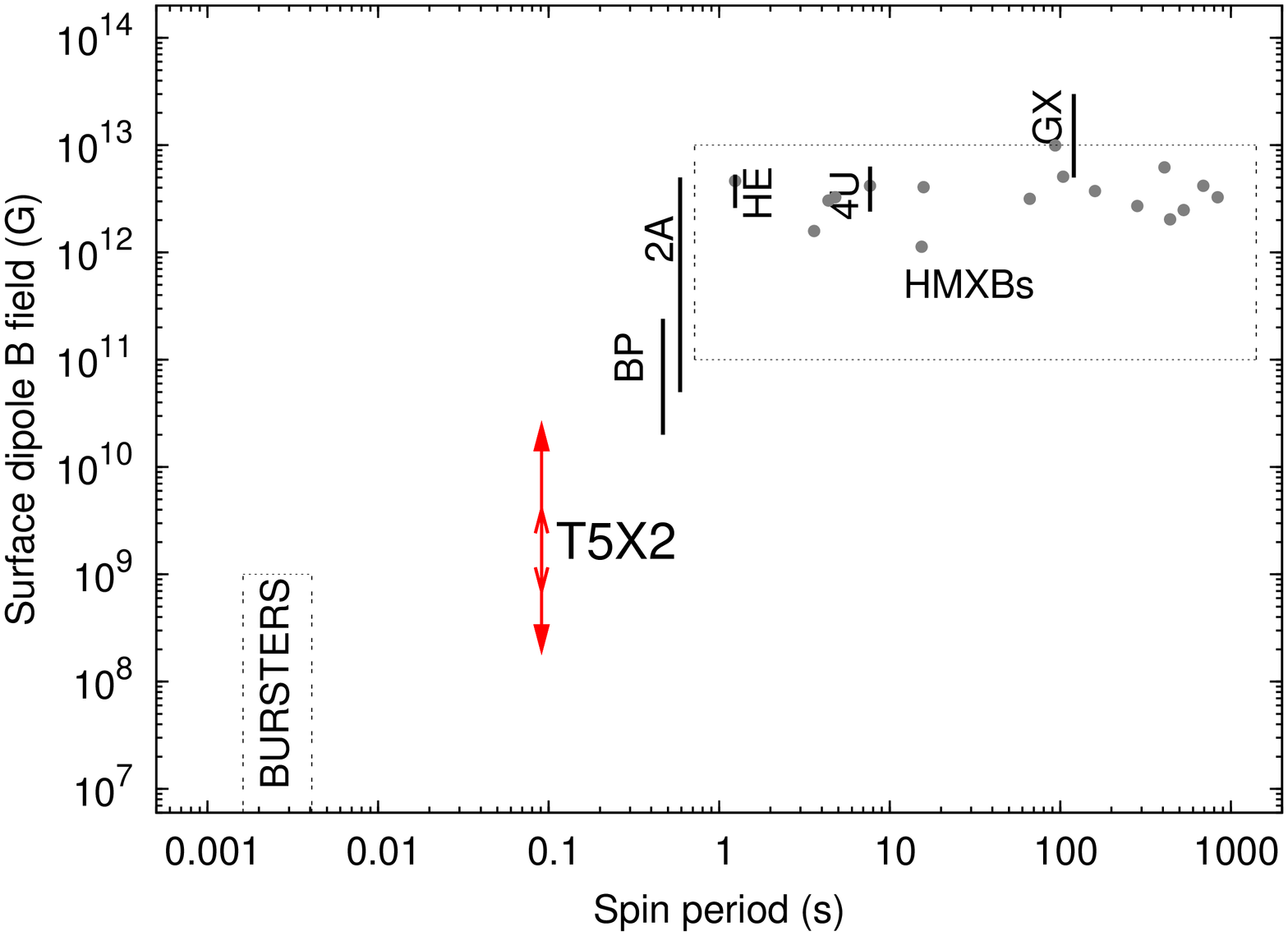}}}
  \caption{Magnetic field strength vs. spin period in accreting
NSs. The dotted rectangle in the lower left corner (labeled
'bursters') shows the location of all previously know thermonuclear
burst sources with measured spin
\citep[][and references therein]{Psaltis99c,Chakrabarty03,Patruno10}.
The dotted rectangle in the upper right corner (labeled 'HMXBs') shows
the location of X-ray pulsars in high-mass X-ray binaries
(\citealt{Bildsten97}; filled gray circles show X-ray pulsars with $B$
measured from cyclotron lines, \citealt{Caballero11}). Red double-head
arrows show the location of T5X2 \citep[using the two magnetic field
estimates from][]{Papitto11,Miller11}. Vertical black lines mark the
location of other slow pulsars in LMXBs ($P_\mathrm{s}$$>$0.01~s),
none of which has shown thermonuclear bursts to date. These are
labeled with the first two characters of the source names, which
follow:
2A~1822--371
\citep
%
%[$P_\mathrm{s}$=0.59~s; $P_\mathrm{o}$=5.6~hr;
%B=5$\times$10$^{10}$--5$\times$10$^{12}$~G;]
%[B fromL-spinup]
%
{Jonker01},
4U~1626--67 
\citep
%
%[$P_\mathrm{s}$=7.66~s; $P_\mathrm{o}$=0.7~hr; B=2.4--6.3$\times$10$^{12}$~G;]
%[Bfrom cyclotron and L-QPO]
%
{Rappaport77,Orlandini98},
GRO~1744--28
\citep[the ``bursting pulsar'', BP;][]
%
%[$P_\mathrm{s}$=0.47~s; $P_\mathrm{o}$=284.0~hr;
%B=2$\times$10$^{10}$--2.4$\times$10$^{11}$~G;]
%[B from L-propeller and L-QPO, respectively]
%
{Kouveliotou96,Finger96,Cui97,Sturner96},
Her~X-1 
\citep
%
%[$P_\mathrm{s}$=1.24~s; $P_\mathrm{o}$=40.8~hr; B=2.6--5.3$\times$10$^{12}$~G;]
%[B from cyclotron]
%
{Truemper78,Bildsten97,Mihara90,Fiume98}
and GX~1+4 
\citep
%
%[$P_\mathrm{s}$=120~s; $P_\mathrm{o}$=27864~hr;
%B=5$\times$10$^{12}$--3$\times$10$^{13}$~G;]
%[B from L-spindown and L-propeller; $P_\mathrm{o}$ from Pereira99]
%
{Chakrabarty97b,Hinkle06,Cui97}.
}
    \label{fig:BvS}
 \end{center}
\end{figure}

%\columnbreak

\section{Discussion}
\label{sec:discussion}
The bursting behavior of T5X2 strikes as surprising for several
reasons. 
During the last three decades bursts at persistent luminosities
$L_\mathrm{pers}$$\gtrsim$0.2~$L_\mathrm{Edd}$ had proven exceptional
and extremely difficult to detect, even when studying a large sample
of bursters \citep[e.g.][and references
therein]{Cornelisse03,Galloway08}. This decrease of burst rate at high
$L_\mathrm{pers}$, opposite to what standard burst theory predicts
\citep{Fujimoto81,Bildsten98}, has been attributed in the literature
to several effects, including: i) stable thermonuclear burning
becoming more important at high $\dot{m}$ and consuming an increasing
fraction of the accreted fuel \citep{Paradijs88}; ii) deflagration
fronts or ``flames'' propagating on the NS surface and consuming part
of the fuel \citep{Bildsten95} and iii) non-spherical accretion confined
to a fraction of the NS surface that would increase with
$L_\mathrm{pers}$ \citep{Bildsten00}.
The large number of bursts observed from T5X2 in October--November
2010 (at an average burst rate over more than a month of
4.9~hr$^{-1}$; Sec.~\ref{sec:results}) make T5X2 the most prolific
source of thermonuclear bursts known to date. Notably, such copious
burst activity was observed when $L_\mathrm{pers}$ was in the range
0.1--0.5~$L_\mathrm{Edd}$, a regime where burst rate is found to
decrease drastically in all other bursters. The only source that has
shown (albeit sporadically) such high burst rates at similar
$L_\mathrm{pers}$ is Cir~X-1: on May 2010 it featured burst recurrence
times as short as 1000--2000~s, when $L_\mathrm{pers}$ was
$\sim$0.2~$L_\mathrm{Edd}$ \citep[][]{Linares10d}. Two other well
known sources of ``high-$\dot{m}$'' bursts, GX~17+2 and Cyg~X-2, have
shown bursts much less frequently \citep[mean burst rate over more
than 10~yr was 0.05 and 0.1~hr$^{-1}$, respectively;][]{Galloway08}
and when $L_\mathrm{pers}$ was close to $L_\mathrm{Edd}$. Other
systems have shown thermonuclear bursts at intermediate accretion
rates ($L_\mathrm{pers}$$\simeq$0.1--0.5~$L_\mathrm{Edd}$; including
GX~3+1, GX~13+1, Ser~X-1, 4U~1636--53, 4U~1735-44, 4U~1746-37), but
only sporadically \citep[$t_\mathrm{wait}$ of at least one hour and
typically much longer;][and references therein]{Galloway08}.
T5X2 is therefore exceptional in that it follows the expected burst
rate at high $\dot{m}$ \citep{Fujimoto81,Bildsten98} much more closely
than any other known burster \citep{Cornelisse03,Galloway08}. We
extend on this comparison between thermonuclear burning theory and
T5X2 in Section~\ref{sec:trec}.

Moreover, T5X2 is the first slow X-ray pulsar (spin period
$P_\mathrm{s}$$>$10~ms) to show thermonuclear bursts. None of the
slower ($P_\mathrm{s}$$\gtrsim$1~s) ``classical'' X-ray pulsars in
HMXBs have shown thermonuclear bursts to date, even if their
persistent luminosity varies over the same range as in bursters. This
is usually attributed to the stronger dipolar NS magnetic field ($B$)
in HMXBs channeling the accretion flow into a much smaller area than
LMXBs, leading to stable burning of all the accreted fuel due to a
very high local $\dot{m}$ (\citealt{Joss80}; see also
Sec.~\ref{sec:bfield}). On the other hand, only 7 out of the more than
90 bursters known have shown X-ray pulsations in the persistent
emission (i.e., 7 out of the 14 accreting millisecond pulsars have
shown bursts to date). Before the discovery of T5X2 the slowest
spinning burster had a spin frequency more than 20 times higher than
T5X2 \citep[IGR~J17511--3057, with a spin frequency
$\nu_\mathrm{s}$=245~Hz;][]{Markwardt09,Altamirano10c}. T5X2 therefore
bridges the gap between ``pulsars that don't burst and bursters that
don't (typically) pulse'' \citep[][and references
therein]{Bildsten98}. We show this in Figure~\ref{fig:BvS} by
comparing both the $B$ and $P_\mathrm{s}$ values of T5X2 to those of
LMXBs (including a few peculiar LMXBs) and HMXBs. Figure~\ref{fig:BvS}
clearly shows that T5X2 features values of $B$ and $P_\mathrm{s}$
intermediate between bursters and HMXBs. We discuss the consequences
of such a high $B$ and long $P_\mathrm{s}$ for thermonuclear burst
regimes in Sections~\ref{sec:bfield} and \ref{sec:rotation},
respectively.

\begin{figure}[h!]
  \begin{center}
  \resizebox{1.0\columnwidth}{!}{\rotatebox{-0}{\includegraphics[]{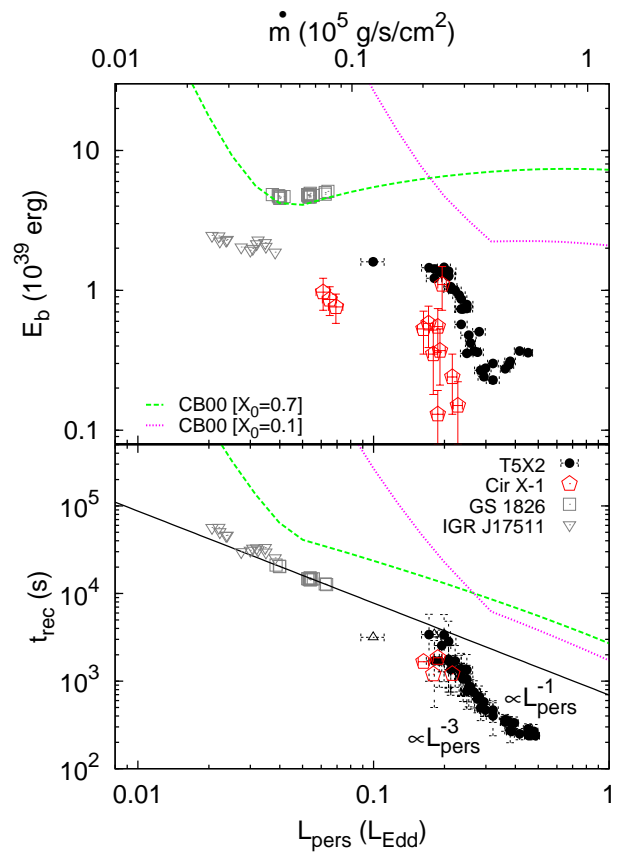}}}
  \caption{Burst energy {\it(top panel)} and recurrence time
{\it(bottom panel)} vs. persistent luminosity (bottom axis) and
inferred mass accretion rate per unit surface (top axis) for four
bursters, as indicated on the bottom panel: T5X2 (filled black
circles; this work), GS~1826--24 \citep[``the clocked burster'', open
gray squares; from][]{Galloway04}, Cir~X-1 \citep[open red pentagons;
from][]{Linares10d} and IGR~J17511--3057 \citep[open gray triangles;
from][]{Falanga11}. The solid line shows the empiric
$t_\mathrm{rec}$~$\propto$~$L_\mathrm{pers}^{-1.05}$ relation found
for GS~1826--24 by \citet[][]{Galloway04}, similar to the
$t_\mathrm{rec} \propto L_\mathrm{pers}^{-1.1}$ relation found in
IGR~J17511--3057 \citep{Falanga11}.
We also show the $E_\mathrm{b}$-$\dot{m}$ and
$t_\mathrm{rec}$-$\dot{m}$ relations predicted by two ignition
models from \citet[][labeled CB00]{CB00} with different accreted
hydrogen fractions: $X_0$=0.7 (green dashed line) and $X_0$=0.1
(magenta dotted line). In both models the metallicity was assumed to
be $Z$=0.02 and the heat flux from the crust was fixed at
$Q_\mathrm{b}$=0.1~MeV~nucleon$^{-1}$.}
    \label{fig:trec}
%\epsscale{1.0}
 \end{center}
\end{figure}

The shortest burst wait times known to date \citep[][5.4~and 3.8~min,
  respectively]{Linares09c,Keek10} were based on sets of typically two
  or three consecutive bursts followed by much longer periods without
  bursts. The bursts from T5X2 presented herein are remarkably
  quasi-periodic, not only in the mHz QPO phase, as witnessed by the
  well defined and smoothly evolving wait time in individual bursts
  ($t_\mathrm{wait}$; see Figs.~\ref{fig:evol} \& \ref{fig:ob}). This
  behavior is analogous to the ``clocked burster''
  \citep[GS~1826--24;][]{Tanaka89,Ubertini99,Galloway04,Heger07b} and
  IGR~J17511--3057 \citep{Falanga11}, which have shown bursts at
  regular intervals with an approximate
  $t_\mathrm{rec}$~$\propto$~$L_\mathrm{pers}^{-1}$ relation, although
  at lower accretion rates than T5X2. For this reason, we compare in
  Figure~\ref{fig:trec} the burst energies and recurrence times of
  T5X2 to those of these two sources, as well as Cir~X-1. The bursts
  from T5X2 (at $\dot{m} \sim$0.1--0.5~$\dot{m}_\mathrm{Edd}$) are
  more frequent and less energetic than those from GS~1826--24 and
  IGR~J17511--3057 (at $\dot{m}
  \sim$0.02--0.06~$\dot{m}_\mathrm{Edd}$). Figure~\ref{fig:trec} also
  shows that the $t_\mathrm{rec}$~$\propto$~$L_\mathrm{pers}^{-3}$
  relation that we find in regime B is unique to T5X2. The bursts from
  Cir~X-1, however, have again a strong resemblance to those of T5X2,
  with energies in the same range ($\lesssim 10^{39}$~erg) and a
  similar $E_\mathrm{b}$-$L_\mathrm{pers}$ relation (a break or
  transition is also suggested by the Cir~X-1 data;
  Figure~\ref{fig:trec}).

\subsection{Bursting regimes vs. burning regimes:\\ the need for heat}
\label{sec:trec}

In the present Section we compare in detail the T5X2 burst properties
  with theoretical predictions. Models of nuclear burning in the
  envelope of (non-magnetic, non-rotating) accreting NSs predict four
  different burning regimes on a NS accreting a mixture of hydrogen
  (H; mass fraction $X_0$), helium (He; mass fraction $Y_0$) and heavy
  elements \citep[mainly CNO, mass fraction $Z$; see][and references
  therein]{Woosley76,Fujimoto81,Taam81,Bildsten98,CB00}.
At the highest accretion rates,
close to or higher than $\dot{m}_\mathrm{Edd}$,
both H and He burn stably and no bursts are expected (Sec.~\ref{sec:msb}).
For accretion rates 
\begin{equation}
\dot{m} > \dot{m}_\mathrm{sHb} \simeq 0.008~\dot{m}_\mathrm{Edd}\left(\frac{0.7}{X_0}\right)\left(\frac{Z}{0.01}\right)^{1/2},
\label{eq:shb}
\end{equation}
\citep[][where we assumed an opacity of 0.04~cm$^2$~g$^{-1}$ and the
  value of $\dot{m}_\mathrm{Edd}$ given in
  Sec.~\ref{sec:data}]{Bildsten98}, H burns stably between bursts
at a constant rate, via the so-called ($\beta$-limited) ``hot-CNO''
cycle \citep[][and references therein]{Bildsten98,CB00}. In this range
all bursts are triggered when He burning at the base of the accreted
layer becomes thermally unstable (He ignition). The column depth (or
density) at the base of the burning layer when ignition occurs is
known as ignition depth, $y_\mathrm{ign}$, and the time between bursts
is simply $t_\mathrm{rec}=y_\mathrm{ign}/\dot{m}$.

As H burns at a constant rate (temperature and $\dot{m}$ independent,
but proportional to $Z$), the time it takes to consume all H in a
sinking fluid element depends only on $X_0$ and $Z$:
\begin{equation}
t_\mathrm{burn}=27~\mathrm{hr}~\left(\frac{X_0}{0.7}\right)\left(\frac{0.01}{Z}\right)~(1+z)/1.31 
\label{eq:tburn}
\end{equation}
\citep[as seen by an observer; e.g.,][]{Galloway06b}, where
1+$z$=(1-2$GM/Rc^2$)$^{-1/2}$=1.31 is the gravitational redshift on the
surface of a NS with mass $M$=1.4~$M_\odot$ and radius $R$=10~km. 
The longest burst recurrence time that we measure in T5X2 is
$t_\mathrm{rec}\sim$1~hr. For Solar abundances this implies that
$t_\mathrm{rec} << t_\mathrm{burn}$, i.e., there is no time to consume
all H between the T5X2 bursts unless the fuel is substantially H-poor
and/or metal rich (see below).
In general, for accretion rates 
\begin{equation}
\dot{m} > \dot{m}_\mathrm{dep} \simeq~0.04~\dot{m}_\mathrm{Edd}\left(\frac{0.7}{X_0}\right)\left(\frac{Z}{0.01}\right)^{13/18}
\label{eq:mdep}
\end{equation}
\citep[][for $M$=1.4~$M_\odot$ and $R$=10~km]{Bildsten98}, there is no
time to burn the accreted H before reaching ignition conditions (i.e.,
$t_\mathrm{rec} < t_\mathrm{burn}$) so that He ignites in a mixture of
H and He. In this regime $y_\mathrm{ign}$ does not depend sensitively
on $\dot{m}$, and a simple $t_\mathrm{rec} \propto \dot{m}^{-1}$
relation is expected. For $\dot{m} < \dot{m}_\mathrm{dep}$ instead,
there is enough time to deplete all H before the base of the accreted
layer reaches ignition conditions. Bursts are then triggered in the
absence of H (pure He ignition). In this regime $y_\mathrm{ign}$
decreases with increasing $\dot{m}$, which results in a steeper
decline of the burst recurrence time as $\dot{m}$ increases, close to
$t_\mathrm{rec} \propto \dot{m}^{-3}$ \citep{CB00}.

This transition between pure He and mixed H/He ignition
regimes can be seen in the $E_\mathrm{b}$-$\dot{m}$ and
$t_\mathrm{rec}$-$\dot{m}$ relations in Figure~\ref{fig:trec}, where
we show two semi-analytic models from \citet{CB00} with
different compositions. Even though the transition from pure He to
mixed H/He ignition is expected at
$\dot{m}_\mathrm{dep}$$\sim$0.05$\dot{m}_\mathrm{Edd}$ for the case
of Solar abundances, Figure~\ref{fig:trec} and Equation~\ref{eq:mdep}
clearly show that changes in the accreted composition can increase
$\dot{m}_\mathrm{dep}$ by a factor of at least 10. 
In particular, Figure~\ref{fig:trec} shows that ignition models with
low H abundance, [$X_0$=0.1, $Z$=0.02], can reproduce the change in
slope of both the $E_\mathrm{b}$-$L_\mathrm{pers}$ and
$t_\mathrm{rec}$-$L_\mathrm{pers}$ relations that we observe in T5X2
at $L_\mathrm{pers} \simeq 0.3 L_\mathrm{Edd}$, i.e., at the
transition between regimes B and C (Sec.~\ref{sec:burstmdot},
Tables~\ref{table:power} \& \ref{table:regimes}). With such low $X_0$,
H can be depleted before reaching He ignition at accretion rates much
higher than in the case of Solar abundances, which highlights the
importance of fuel composition in the burning regimes.

The average T5X2 accretion-to-burst energy ratio
($\alpha$=$Q_\mathrm{grav}$(1+$z$)/$Q_\mathrm{nuc}$),
$\langle\alpha\rangle = 102$, corresponds to a total nuclear energy
release during the bursts
$Q_\mathrm{nuc}$$\simeq$2.8~MeV~nucleon$^{-1}$ \citep[see
also][]{Motta11}. Taking $Q_\mathrm{nuc} = 1.6+4\langle
X\rangle$~MeV~nucleon$^{-1}$ \citep[which assumes complete burning of
the accumulated fuel and $\sim$35\% neutrino energy loss;][and
references therein]{Galloway08} we find an average H mass fraction
over the burning layer $\langle X\rangle$$\simeq$0.3. Such low
inferred $Q_\mathrm{nuc}$ lends support to the low H fraction fuel
scenario for T5X2 proposed above.
In summary, the slopes of the $E_\mathrm{b}$-$L_\mathrm{pers}$ and
$t_\mathrm{rec}$-$L_\mathrm{pers}$ relations as well as the $\alpha$
values observed during the T5X2 outburst, indicate a low accreted H
fraction and strongly suggest a transition from the pure He
ignition regime to the mixed H/He ignition regime happening during
the outburst rise when $\dot{m}$ increases above
$\sim$0.3~$\dot{m}_\mathrm{Edd}$. It is worth noting that the reverse
transition is observed during the outburst decay when $\dot{m}$ {\it
drops} below 0.3~$\dot{m}_\mathrm{Edd}$ (i.e., the outburst rise and
decay tracks overlap and no hysteresis is seen in
Figures~\ref{fig:lumlum} \& \ref{fig:lumlumlin}).

We stress that the link between the observed T5X2 {\it bursting}
regimes and the theoretical {\it burning} regimes that we put forward
is based on the $t_\mathrm{rec}$-$L_\mathrm{pers}$ and
$E_\mathrm{b}$-$L_\mathrm{pers}$ relations. Modeling and
interpretation of the individual burst light curves is beyond the
scope of this work, yet we note that different burst light curves and
peak luminosities are predicted in the different ignition regimes
\citep[e.g.][]{Woosley04}. Pure He bursts are expected to show faster
rise times than mixed H/He bursts. This agrees qualitatively with the
identification of regime B as pure He ignition: $t_\mathrm{rise}$ is
shorter in regime B than in regime C (Fig.~\ref{fig:alpha}). The large
change in $L_\mathrm{pers}$, however, could influence the observed
timescales along the burst rise and decay, as these are measured after
subtracting the persistent emission. It should also be noted that at
low accretion rates (near
$\dot{m}_\mathrm{sHb}~\simeq~0.08~\dot{m}_\mathrm{Edd}$ for $X_0$=0.1,
$Z$=0.02) there should be a transition to H-ignited bursts, not
included in the models discussed in this Section \citep{CB00}.

If the link between bursting and burning regimes that we propose is
correct, to our knowledge this is the first time that the transition
between pure He and mixed H/He ignition is observed in a single
source.
There is, however, a systematic and interesting discrepancy evident in
Figure~\ref{fig:trec}. Even when including compressional heating
or when increasing the base heat flux from the NS crust \citep[up to 2
MeV~nucleon$^{-1}$; fixed at 0.1~MeV~nucleon$^{-1}$ in
Fig.~\ref{fig:trec};][]{CB00}, ignition models predict higher burst
energies and longer recurrence times than those we find in T5X2, by a
factor close to 10 in most cases (i.e., larger than the distance
uncertainty on $L_\mathrm{pers}$ and $E_\mathrm{b}$,
Sec.~\ref{sec:data}).
This large difference between the observed and predicted values
of $E_\mathrm{b}$ and $t_\mathrm{rec}$, together with the lack of
detailed modeling of the T5X2 burst light curves, prevents a
conclusive identification of the observed bursting regimes.

We propose that such discrepancy could be explained by the presence of
an extra source of heat in the NS envelope not accounted for by
ignition models (which typically consider only hot-CNO
heating). Additional heat would act to reduce $y_\mathrm{ign}$ and
thereby decrease $E_\mathrm{b}$ and $t_\mathrm{rec}$, explaining the
T5X2 observations presented herein. Several interesting possibilities
for the nature of this extra source of heat have been partially
investigated, including: i) heating due to stable He burning
\citep[triple alpha reaction rates have been recently debated;
e.g.,][]{Ogata09,Dotter09,Peng10}; ii) heating due to deep burning of residual
H \citep[e.g.,][who already pointed out that it can lead to weaker and
more frequent bursts than expected]{Taam96}; iii) thermal inertia,
``hot ashes'' or heat from previous bursts, which could become
important for T5X2 as it features $t_\mathrm{rec} \sim 200s$, the
shortest recurrence times between thermonuclear bursts ever observed
\citep[time-dependent simulations of a series of bursts are needed to
investigate this in detail; e.g.,][]{Heger07} and iv) turbulent
friction at the base of the spreading layer, which could release
substantial amounts of heat \citep{Inogamov10}.
It is worth noting that an independent study of the same source
\citep{Degenaar11c} reached similar conclusions, suggesting the
presence of a ``shallow heat source'' in T5X2 in order to reconcile
quiescent observations and crust heating/cooling theory.
Without regard to the exact nature of this heat source, we have shown
in the present work that the burst properties of T5X2 place new
constraints on the thermal properties of a fast-accreting and
frequently-bursting NS.

%average alpha: 

%gs1826 = 42 \citep{Galloway04}

%cirx-1 = 150 \citep{Linares10d}

%T5X2 = 102 (this work)

\subsection{mHz QPOs and marginally stable burning}
\label{sec:msb}

The mHz QPOs from T5X2 \citep[Sec.~\ref{sec:mhzqpo}; see
also][]{Linares10c} have distinctive properties that clearly set them
apart from the previously known mHz QPOs
\citep[Sec.~\ref{sec:intro};][]{Revnivtsev01,Altamirano08d}. The
persistent luminosity that we measure in T5X2 when mHz QPOs are
present is about 10 times higher than that observed in previous mHz
QPO sources ($L_\mathrm{pers}$ higher by a factor of 4--25 taking into
account the observed ranges: 0.02--0.1~$L_\mathrm{Edd}$ in atoll
sources as opposed to 0.4--0.5~$L_\mathrm{Edd}$ in T5X2;
Table~\ref{table:mhzqpo}). For this reason we refer to the previously
known mHz QPOs as ``low-$L_\mathrm{pers}$ mHz QPOs''.

Bright bursts and low-$L_\mathrm{pers}$ mHz QPOs alternate, while
remaining clearly distinguishable. Instead, in T5X2 bursts smoothly
evolve into mHz QPOs and vice versa (Sec.~\ref{sec:mhzqpo}), a
phenomenon never observed before. Strikingly, the same qualitative
evolution from bright infrequent bursts to faint and frequent bursts,
mHz QPOs and ultimately stable burning is predicted to happen as
$\dot{m}$ increases by both one-zone models and detailed simulations
of nuclear burning on NSs accreting near the boundary between unstable
and stable He burning \citep{Heger07}. 
We can therefore identify with confidence the mHz QPOs from T5X2 with
marginally stable burning on the NS surface. The evidence that links
low-$L_\mathrm{pers}$ mHz QPOs with marginally stable burning is less
conclusive, but remains valid \citep[see][for
details]{Revnivtsev01,Yu02,Altamirano08d}.

Analytic estimates place the
threshold for stable He burning at
\begin{equation}
\dot{m}_\mathrm{sb} \simeq 1.1~\dot{m}_\mathrm{Edd}\left(\frac{1.7}{1+X_0}\right)^{3/4}\left(\frac{Y_0~\mu}{0.3\times0.6}\right)^{1/2},
\label{eq:sb}
\end{equation}
where $\mu$ is the mean molecular weight of the accreted fuel
\citep[][assuming again $M$=1.4~$M_\odot$ and $R$=10~km]{Bildsten98}.

The persistent luminosity where mHz QPOs are observed in T5X2
(0.4--0.5~$L_\mathrm{Edd}$) is therefore closer to the expected value
of $\dot{m}_\mathrm{sb}$ than what was seen in low-$L_\mathrm{pers}$
mHz QPOs, yet still inconsistent with the value predicted by theory,
which is higher by a factor of $\sim$2 (unless Terzan 5 is at
$\sim$9~kpc instead of 6.3kpc, Sec.~\ref{sec:data}).
Invoking H-poor fuel makes this
discrepancy even larger, as it can increase the expected
$\dot{m}_\mathrm{sb}$ by a factor 2 (for $X_0$=0.1, $Z$=0.01;
Eq.~\ref{eq:sb}). A more massive NS can have unstable burning at
higher $\dot{m}$ than a less massive star, but given the weak
(M$^{1/2}$) scaling of $\dot{m}_\mathrm{sb}$ different NS masses
cannot explain the large difference (factor 4--25) in
$L_\mathrm{pers}$ between T5X2 and the other known sources of mHz
QPOs.
Another remarkable difference betwen the mHz QPOs presented herein and
the low-$L_\mathrm{pers}$ mHz QPOs resides in their fractional rms
spectrum (Sec.~\ref{sec:data}): we find a fractional rms amplitude
that increases with energy between 2 and 20~keV
(Sec.~\ref{sec:mhzqpo}), while \citet{Revnivtsev01} reported
fractional amplitudes decreasing with increasing photon energy between
2 and 5 keV. The energy-averaged fractional amplitudes are, however,
similar \citep[1.3--2.2 \% in T5X2, this work; 0.7--1.9 \% in
low-$L_\mathrm{pers}$ mHz QPOs,][]{Revnivtsev01}.
The increase of fractional amplitude with energy that we find is
shallower than the linear increase predicted by models of temperature
oscillations at the NS surface \citep[developed in the context of
burst oscillations;][]{Piro06}.

An interesting possibility is that the low-$L_\mathrm{pers}$ mHz QPOs
trace the boundary of stable H burning (Eq.~\ref{eq:shb}), whereas the
mHz QPOs that we discovered in T5X2, at a higher accretion rate, occur
at the boundary of stable He burning (Eq.~\ref{eq:sb}). This remains
speculative at present given the lack of published theoretical
work on this particular topic. A specific analysis of oscillatory
burning at the thermal stability boundary of H burning is needed to
address this hypothesis. The oscillatory behavior at the marginally
stable point is generic \citep{Paczynski83} and it could also occur
when H burning stabilizes. If low-$L_\mathrm{pers}$ mHz QPOs do happen
at the H burning stability boundary, the need to invoke confined
accretion to explain their low $L_\mathrm{pers}$ would vanish.

\citet{Heger07} found that the QPO frequency is mainly sensitive to
the accreted H fraction ($X_0$) and the NS surface gravity: increasing
the surface gravity or decreasing $X_0$ leads to higher mHz QPO
frequencies. The highest $\nu_\mathrm{QPO}$ values that we find in
T5X2 (4.2~mHz) are about a factor of 3 lower than those of the
low-$L_\mathrm{pers}$ mHz QPOs
\citep[Table~\ref{table:mhzqpo};][]{Revnivtsev01,Altamirano08d}. If
the accreted fuel has a similar composition, and if the mHz QPOs have
the same origin, this would suggest a less compact NS in T5X2 than in
the low-$L_\mathrm{pers}$ mHz QPO sources (the ``atoll'' sources
4U~1636-536, 4U~1608-52 and Aql~X-1). Comparing to models of
marginally stable burning, we find that if X$_0$$<$0.7 the surface
gravity of the NS in T5X2 must be $g$ =
$GM/R^2$$<$1.9$\times$10$^{14}$~cm~s$^{-2}$ \citep[Figure~9
in][]{Heger07}. Even though it is subject to theoretical (based on
analytic one-zone model) and observational (a new outburst of T5X2
could show higher $\nu_\mathrm{QPO}$) uncertainties, this illustrates
how a comparison between marginally stable burning models and mHz QPO
properties can be used to constrain the NS compactness.

Despite the striking similarities between the T5X2 burst properties
and the general bursting behavior predicted by models of nuclear
burning near the transition from unstable to stable burning
\citep[cf. Fig.~\ref{fig:evol} in this work and Fig.~5 in][]{Heger07},
interesting differences remain. First, as explained above, the
$\dot{m}$ where marginally stable burning is expected
\citep[0.925~$\dot{m}_\mathrm{Edd}$][]{Heger07} is higher than the
highest inferred $\dot{m}$ where we observe mHz QPOs in T5X2
(0.5~$\dot{m}_\mathrm{Edd}$). Second, \citet{Heger07} find a sharp
transition from bursts to mHz QPOs and finally stable burning
(occurring between 0.923--0.95~$\dot{m}_\mathrm{Edd}$), while we
observe in T5X2 a smooth evolution from bursts to mHz QPOs, and
vice versa (between 0.1--0.5~$\dot{m}_\mathrm{Edd}$;
Sec.~\ref{sec:results}). Time-dependent simulations of nuclear burning
tailored to T5X2 would be of high interest, in particular exploring
the H-poor fuel range. Third, we find that the disappearance of the
mHz QPOs in T5X2 is not linked to an increase in $\dot{m}$ but to an
actual {\it drop} of $L_\mathrm{pers}$ that happens when the accretion
state changes (from horizontal to normal \& flaring branches). This
suggests that the geometry/configuration of the accretion flow plays a
role in setting the nuclear burning stability boundary.

\subsection{The role of an intermediate magnetic field}
\label{sec:bfield}

While most models of thermonuclear bursts on accreting NSs assume that
the NS magnetic field is negligible
\citep[][]{Fujimoto81,Taam82,Paczynski83,Woosley04}, \citet{Joss80}
showed that the presence of a strong ($B$$\gtrsim$10$^{12}$~G) magnetic
field can affect the stability of nuclear burning in different ways. A
strong magnetic field reduces the (conductive and radiative)
opacities, allowing for more efficient heat transport and thereby
stabilizing burning. Applying disk-magnetosphere interaction models to
T5X2, \citet{Papitto11} estimated
$B$=2$\times$10$^8$--2.4$\times$10$^{10}$~G from the luminosity range at
which pulsations were detected, whereas \citet{Miller11} further
constrained $B$=(0.7--4)$\times$10$^9$~G from the Fe line profile. The
fact that T5X2 shows thermonuclear bursts suggests that the magnetic
field needed to suppress the thermal instability and quench
thermonuclear bursts must be greater than $\sim$10$^{10}$~G, in
accordance with theoretical expectations \citep{Joss80}.

A strong magnetic field also affects convective heat transport and
mixing \citep[e.g.][]{Joss80}. \citet{Bildsten95} proposed that even
in cases where thermonuclear burning is unstable the presence of a
strong magnetic field can suppress convection in the NS envelope,
stalling the propagation of the burning front (diffusive heat
transport being slower than convection) and preventing the fast ignition
that causes type I X-ray bursts \citep[see also][]{Bildsten98}. Again,
the mere presence of type I X-ray bursts in T5X2 strongly suggests
that the magnetic field required to suppress convective burning fronts
must be higher than $\sim$10$^{10}$~G, in agreement with analytical
estimates \citep{Joss80,Bildsten97b}. Rise times are rather long
though (Fig.~\ref{fig:alpha}), which could perhaps indicate that $B$ is
strong enough to slow down the burning fronts \citep{Bildsten97b}. In
summary, the bursting properties of T5X2 presented herein (in
particular combined with refined measurements of its magnetic field
strength) can place new constraints on the physics of convection and
heat transport under intermediate magnetic fields
(10$^8$--10$^{10}$~G).

Finally, a 10$^8$--10$^{10}$~G magnetic field can also channel the
accretion flow onto the magnetic polar caps, as suggested by
theoretical models of disk-magnetosphere interaction \citep{Lamb73}
and by the presence of X-ray pulsations in T5X2
\citep{Strohmayer10,Papitto11}. Accretion confined to the magnetic
polar caps may break the spherical symmetry assumed by most
thermonuclear burst models, and increase $\dot{m}$ for a given
(observed) $L_\mathrm{pers}$ (Sec.~\ref{sec:data}). The impact on the
burning properties depends on the size of the polar caps and on how
soon the accreted fuel spreads laterally while sinking into the NS
envelope under the influence of a magnetic field. \citet{Bildsten97b}
found that for dipolar magnetic fields $B$$<$ (2--4)$\times 10^{10}~G$,
the accreted fuel spreads before igniting and the spherically
symmetric case is recovered. This suggests that the T5X2 burst
properties can still be compared to spherically symmetric ignition
models (Sec.~\ref{sec:trec}).

As the size and ``depth'' of the polar caps are nevertheless
ill-constrained quantities, we explore a simple scenario in which the
fuel in T5X2 is confined into a 10\% of the NS surface down to
ignition depth, $y_\mathrm{ign}$. From geometrical considerations,
assuming the same $y_\mathrm{ign}$ predicted by ignition models
\citep{CB00}, in this confined scenario bursts would be 10 times less
energetic and 10 times more frequent than in the spherically symmetric
case, which would explain the mismatch between predicted and observed
$E_\mathrm{b}$ and $t_\mathrm{rec}$ (Sec.~\ref{sec:trec}). However,
the accretion rate per unit area would also increase by a factor 10,
which for the observed $L_\mathrm{pers}$ implies that unstable burning
would operate at $\dot{m}$ as high as $\sim$5 times
$\dot{m}_\mathrm{Edd}$. This is well above the predicted threshold for
stable steady burning of H and He \citep[$\dot{m}_\mathrm{sb}$,
Eq.~\ref{eq:sb}; see also][]{Bildsten98}, and we therefore consider
this scenario unlikely. If the accreted material contains no H,
however, $\dot{m}_\mathrm{sb}$ can be several times larger than
$\dot{m}_\mathrm{Edd}$ \citep{Bildsten98}.

\subsection{Rotation, mixing and burning}
\label{sec:rotation}

The unique bursting behavior in T5X2 is, in essence, much closer to
what theory predicts than any other burster known to date. Theoretical
burning regimes were mostly based on models that assume a non-rotating
NS and radial infall of the accreted mass
\citep{Fujimoto81,Bildsten98}. These are admittedly simplistic
assumptions, as NSs in LMXBs can have spin frequencies in excess of
600~Hz and the orbital frequencies in the innermost accretion disk are
well above 1000~Hz. The accretion of matter with angular momentum onto
a spinning NS introduces a shear in its surface layers \citep[][and
references therein]{Piro07,Keek09}. The main effect of rotation which
may have an impact on burning regimes is turbulent mixing. In
particular, \citet{Keek09} found that such rotationally induced
turbulent mixing stabilizes He burning, decreasing the
burning stability boundary ($\dot{m}_\mathrm{sb}$).

Motivated by the stronger magnetic field (factor $\sim$10) and
slower spin (factor $\gtrsim$20) of T5X2 compared to the rest of
bursters, we speculate that rotationally induced turbulent mixing is
what sets T5X2 apart from the rest of thermonuclear burst sources. The
stronger $B$ field could channel a large fraction of the accreted
matter into the NS magnetic poles and produce a more radial inflow
which would then spread over the whole surface, minimizing shear and
turbulent mixing. Therefore T5X2 seems to meet to a greater extent the
assumptions of zero spin and radial infall mentioned above, which can
explain the better agreement with theory. This in turn suggests that,
as argued by \citet{Piro07} and \citet{Keek09}, the effects of
rotation must be considered to explain the behavior of most
bursters.\\

\section{Summary and Conclusions}
\label{sec:conclusions}

We have presented the discovery of mHz QPOs from the NS-LMXB and 11~Hz
X-ray pulsar T5X2, as well as the full bursting properties during its
2010 outburst, when the persistent luminosity varied by about a factor
of 5 (between
$L_\mathrm{pers}$$\simeq$0.1--0.5~$L_\mathrm{Edd}$). T5X2 showed
copious thermonuclear bursts when
$L_\mathrm{pers}$$\gtrsim$0.2~$L_\mathrm{Edd}$, a regime where
thermonuclear bursts had proven exceptional to date. Burst energies and
recurrence times gradually decreased as $L_\mathrm{pers}$ increased,
turning into a rapid series of faint bursts and smoothly evolving into
a mHz QPO near the outburst peak. Only at the highest
$L_\mathrm{pers}$ range bursts became undetectable during a few short
intervals. This behavior is both unprecedented among bursters and
remarkably similar to the nuclear burning regimes expected on an
accreting NS near the boundary of stable He burning.

We find four different bursting regimes when studying the relation
between burst properties and $L_\mathrm{pers}$ and show that in one of
such regimes the burst recurrence time decays steeply with increasing
$L_\mathrm{pers}$, close to
$t_\mathrm{rec}$~$\propto$~$L_\mathrm{pers}^{-3}$. By confronting the
change in burst properties (in particular, the relation between
$E_\mathrm{b}$, $t_\mathrm{rec}$ and $L_\mathrm{pers}$) to ignition
model predictions, we find evidence of a transition between pure He
and mixed H/He ignition occurring in T5X2 when
$L_\mathrm{pers}$$\simeq$0.3~$L_\mathrm{Edd}$. We note that large
discrepancies remain between the observed T5X2 burst properties and
those predicted by theory at high mass accretion rates. We further
argue that the accreted fuel is H-poor and suggest that an additional
source of heat in the NS envelope is needed to reconcile the observed
and predicted burst properties.

We examine the properties of the mHz QPOs from T5X2 in the context of
marginally stable burning models, and compare them to those of
previously known mHz QPO sources. T5X2 features mHz QPOs with lower
frequencies (by a factor $\sim$3), when $L_\mathrm{pers}$ is
substantially higher (by a factor $\sim$4--25). Finally, we discuss the
role of magnetic field and spin in setting the unique T5X2 burst and
mHz QPO behavior, and speculate that the absence of rotation
effects such as turbulent mixing of the accreted fuel may set T5X2
apart from the rest of bursters.\\

%\textbf{Acknowledgments:} 
We thank M. van der Klis for detailed comments on the manuscript. We
are grateful to the International Space Science Institute in Bern,
where part of this work was completed. ML acknowledges support from
the NWO Rubicon fellowship.

%\bibliographystyle{apj}
%\bibliography{/home/linares/papers/biblio.bib}

\end{document}